\documentclass{aa}
\usepackage{graphicx}
\usepackage{verbatim}\usepackage{verbatim}
\usepackage{verbatim}
\usepackage{txfonts}
\usepackage{stfloats}
\usepackage{color}
\usepackage{amsbsy}
\usepackage{gensymb}
\usepackage{textcomp}
\usepackage{dcolumn}
\usepackage{natbib}
\usepackage{multirow}
\usepackage{longtable}
\usepackage{xspace}
\usepackage{enumitem}
\newcolumntype{d}[1]{D{.}{\cdot}{#1}}
\definecolor{mygray}{gray}{0.6}
\newcolumntype{.}{D{.}{.}{-1}}
\bibpunct[; ]{(}{)}{,}{a}{}{;} 

\newcommand{\lsun}{L$_\odot$}
\newcommand{\Lsun}{L$_\odot$}
\newcommand{\msun}{M$_\odot$}
\newcommand{\Msun}{M$_\odot$}

\newcommand{\vlsr}{$\varv_\mathrm{lsr}$}

\newcommand{\mum}{$\mu$m }
\newcommand{\micron}{$\mu$m }
\newcommand{\kms}{km\,s$^{-1}$}

\newcommand{\hi}{H\textsc{i} }
\newcommand{\Hi}{H\textsc{i} }

\newcommand{\hii}{H\textsc{ii} }

\newcommand{\elll}{$\ell$}
\newcommand{\simm}{$\sim$}

\newcommand{\papertwo}{Paper\,\textsc{II}\xspace}
\usepackage{siunitx}
\newcommand{\sedsLbolMean}{$3.6\times10^{2}$}

\newcommand{\sedsLbolMin}{$8.0\times10^{-2}$}
\newcommand{\sedsLbolMax}{$2.4\times10^{4}$}

\newcommand{\sedsMMin}{$2.3\times10^{-2}$}
\newcommand{\sedsMMax}{$1.7\times10^{3}$}

\newcommand{\sedsRfwhmMean}{0.25}

\newcommand{\sedsRfwhmMin}{0.03}
\newcommand{\sedsRfwhmMax}{1.78}

\newcommand{\sedsNHtwopeakMean}{$3.8\times10^{21}$}

\newcommand{\sedsNHtwopeakMin}{$6.3\times10^{19}$}
\newcommand{\sedsNHtwopeakMax}{$4.1\times10^{22}$}

\newcommand{\sedsTMean}{17.29}

\newcommand{\sedsTMin}{9.65}
\newcommand{\sedsTMax}{41.31}

\newcommand{\sedsTauMin}{$1.9\times10^{-6}$}
\newcommand{\sedsTauMax}{$2.9\times10^{-3}$}

\newcommand{\coRgalMin}{8.50}

\newcommand{\coRhelMin}{0.40}

\newcommand{\agalRfwhmMean}{$3.2\times10^{-1}$}

\newcommand{\varTotalSources}{830}
\newcommand{\varSourcesNoEmission}{37}
\newcommand{\varSourcesWithEmission}{759}
\newcommand{\varSourcesWithEmissionFraction}{91}

\newcommand{\varNumBadCal}{23}
\newcommand{\varNumGoodCalComponents}{1383}
\newcommand{\varNumGoodCalComplexes}{1248}

\newcommand{\varNumBadCalComplexes}{29}
\newcommand{\varNumPos}{1090}
\newcommand{\varNumLosOffComplex}{242}

\newcommand{\varNumComponents}{1415}

\newcommand{\varNumComponentsOn}{1102}

\newcommand{\varNumComplexesOn}{982}

\newcommand{\varNumPosOff}{331}

\newcommand{\varNumPosContaminatedFraction}{39.9}

\newcommand{\varNumComplexesOff}{295}
\newcommand{\varNumComplexesOffFraction}{23.1}
\newcommand{\varIntegrationTime}{1--3}
\newcommand{\varNumComplexes}{1277}
\newcommand{\varNumComplexesPerLoS}{1.2}
\newcommand{\varNumLosSingleComplexes}{966}
\newcommand{\varNumLosSingleComplexesFraction}{88.6}
\newcommand{\varNumLosMultiComplexes}{87}
\newcommand{\varNumLosMultiComplexesFraction}{11.4}

\newcommand{\varNumSEDsFitted}{611}

\newcommand{\varNumSEDsNotFitted}{180}

\begin{document} 
   \title{A New Search for Star Forming Regions \\in the Southern Outer Galaxy}
   \titlerunning{Star Formation in the Outer Galaxy}
   \subtitle{}
   \author{C. K\"onig\inst{1}
			\and
			J. S. Urquhart\inst{2}
			\and
			F. Wyrowski\inst{1}
			\and
			D. Colombo\inst{1}
			\and
			K. M. Menten\inst{1}
          }
   \authorrunning{C. K\"onig et al.}
   \institute{Max-Planck-Institut f\"ur Radioastronomie (MPIfR), 
              Auf dem H\"ugel 69, 53121 Bonn, Germany\\
              \email{koenig@mpifr-bonn.mpg.de}
         \and
         	  School of Physical Sciences, University of Kent, 
              Ingram Building, Canterbury, Kent CT2 7NH, UK
             }
   \date{Received MMMM dd, YYYY; accepted MMMM dd, YYYY}
  \abstract
   {Star-formation in the outer Galaxy is thought to be different from the inner Galaxy, as it is subject to different environmental parameters such as metallicity, interstellar radiation field, or mass surface density that all change with Galactocentric radius. Extending our knowledge on star-formation from the inner to the outer Galaxy helps us to understand the influences of the change of the environment on star formation throughout the Milky Way.}
   {We therefore aimed at getting a more detailed view on the structure of the outer Galaxy, determining physical properties for a large number of star forming clumps and understanding star-formation outside the Solar circle. As one of the largest expanding Galactic supershells is present in the observed region, a unique opportunity is used here to investigate the influence of such an expanding structure on star-formation as well.}
   {We use pointed $^{12}$CO(2--1) observations conducted with the APEX telescope to determine the velocity components towards \varTotalSources\ dust clumps identified from 250\,\mum\ Herschel/Hi-GAL SPIRE emission maps in the outer Galaxy between $225\degree < \ell < 260\degree$. We determined kinematic distances from the velocity components, in order to analyze the structure of the outer Galaxy and to estimate physical properties such as dust temperatures, bolometric luminosities, clump masses, and H$_2$ column densities for \varNumSEDsFitted\ clumps. For this, we determined the dust spectral energy density distributions from archival mid-infrared to sub-millimeter emission maps.}
   {
   We find the identified CO clouds to be strongly correlated with the highest column density parts of the \hi emission distribution, spanning a web of bridges, spurs and blobs of star forming regions between the larger complexes, unveiling the complex three-dimensional structure of the outer Galaxy in unprecedented detail.
   Using the physical properties of the clumps, we find an upper limit of 6\% (40 sources) to be able to form high-mass stars. This is supported by the fact that only 2 methanol Class II masers or 34 known or candidate \hii regions are found in the whole survey area, indicating an even lower fraction to be able to form high-mass stars in the outer Galaxy.
   We fail to find any correlation of the physical parameters of the identified (potential) star forming regions with the expanding supershell, indicating that although the shell organizes the interstellar material into clumps, the properties of the latter are unaffected.
   }
   {Using the APEX telescope in combination with publicly available Hi-GAL, MSX and Wise continuum emission maps, we were able to investigate the structure and properties of a region of the Milky Way in unprecedented detail.}
   \maketitle
\section{Introduction}
Star formation and the processes involved with it, such as disc accretion, molecular outflows, stellar winds, chemical enrichment and energy input into the local environment from radiation, mechanical energy from the outflows, and supernova explosions of the most massive stars, play an important role in determining the structure of the interstellar medium (ISM) and driving the evolution of a galaxy (\citealt{kennicutt2005}). As star formation takes place in molecular clouds, it is not only important to know the distribution of these clouds within the Galaxy but also how different environmental conditions affect their properties, structure and dynamics.
In the Milky Way, the distribution of molecular clouds as traced by CO \citep{Garcia2014,Rice2016,miville2017} shows a strong peak within $\sim$2\,kpc of the Galactic center, and another peak at a distance of $\sim$5\,kpc from the Galactic center, after which the distribution drops off out to \simm20\,kpc. The vast majority of the molecular gas found in the Galaxy is located in the inner Galaxy \citep[i.e. $\sim$85\% within the Solar circle at $R<8.3$\,kpc;][]{miville2017} and as a result most of the previous studies have focused on star formation in the inner part of the Milky Way. Many thousands of low- and high-mass star forming regions have been identified and investigated by a large community \citep[e.g. ][]{Mooney1988,Brand1993,urquhart2014_csc,urquhart2014_rms,Urquhart2018,Elia2017}.
In contrast to the molecular clouds in the inner Galaxy, the clouds located in the outer Galaxy (i.e. outside the Solar circle at $R>8.3$\,kpc) contribute only $\sim$15\% of the total molecular gas of the Milky Way \citep{miville2017}. However, observations towards the outer Galaxy do not suffer from the kinematic distance ambiguities that plague studies of the inner Galaxy \citep{roman2009,Wienen2015} and source confusion is significantly reduced due to a lower density of molecular clouds. There are, therefore, a number of observational advantages to studying molecular clouds located outside the Solar circle. The physical conditions are also very different compared to those found in the inner Galaxy (e.g., the \Hi density decreases, UV radiation field is less intense, and the general cosmic-ray flux and metallicity decreases with increasing Galactocentric distance \citep[e.g.][]{Bloemen1984,Rudolph1997} and so studies of these objects provide valuable insight into how the initial conditions of the clouds, and the physical conditions of their local environment, affect star formation. 
There have been a number of studies that have investigated the distribution and properties of molecular clouds in the outer Galaxy \citep[e.g.][]{Wouterloot1989,May1997,heyer2001,Nakagawa2005,Elia2013a}. These studies found the clouds to be, in general, smaller, less massive and have smaller line-widths compared to clouds located in the inner Galaxy \citep[e.g.][]{Dame1986,solomon1987}. In addition, the conditions of molecular clouds located in different spiral arms might be different \citep[e.g.][]{benjamin2005}. Especially the impact of the entry shock experienced from the material entering a spiral arm should be stronger within the solar circle, as the surface density of the ISM drops by about an order of magnitude around the solar circle \citep{Heyer2015}, and hence the amplitude of the spiral density wave is significantly smaller in the outer Galaxy. Furthermore, as the velocity difference between the ISM and the spiral pattern reverses sign around (and reaches zero at) the co-rotation radius R$_\mathrm{cr}\approx10.9$\,kpc \citep{Koda2016} in the outer Galaxy, the entry shock should be further diminished. Therefore, outside the solar circle radius supernovae explosions may take over as the dominant mechanism determining the state of the ISM \citep[e.g.][]{kobayashi2008}.
In this paper we will build on these previous studies and investigate the properties of a large sample ($\sim$800) of molecular structures located in the Galactic longitude range  $225\degr < \ell < 260\degr$. This region includes a large section of the Perseus and Outer arm and so will allow us to compare the properties of molecular clouds located in the different arms and inter-arm regions \citep[e.g.][]{eden2013,eden2015}. This study has also been designed to complement the recent studies of the inner Galaxy reported by \citet{Urquhart2018}. In an upcoming paper (from here on \papertwo) we will also compare the results of the two studies, allowing us to investigate trends in the clump properties and their distribution over the whole range of Galactocentric distances out to $\sim16$\,kpc.
In this paper, our main goals are to analyze the structure of the Galaxy outside the Solar circle, to identify possible complexes of star formation and characterize and compare their physical properties. We use pointed $^{12}$CO(2--1) observations towards molecular clouds and the rotation curve from \citet{Brand1993} to calculate kinematic distances from the observed radial velocity ($\varv_\mathrm{lsr}$). Archival dust continuum emission maps are used to determine physical properties, following the methods developed in our previous work for the inner Galaxy \citep{Koenig2017,Urquhart2018}.
The paper is structured as follows: in Sect.\,\ref{sect:data} we describe how we selected our sources, the setup for the $^{12}$CO(2--1) observations, how we obtained velocities and determined Galactocentric distances.  In Sect.\,\ref{sec:properties} we present the physical properties obtained from dust spectral energy distributions, using the distances we determined. We discuss these properties and the star forming relations for a detailed look on star formation in the outer Galaxy. In Sect.\,\ref{sec:structures} we will discuss the large-scale structures encountered in this part of the outer Galaxy and investigate the influence of the supershell on its environment. In the final section we give an overview of our findings as well as an outlook on our future work and upcoming papers.
\section{Molecular line data used}
\label{sect:data}
For the present work we identified and selected dust continuum sources from the Herschel Hi-GAL continuum maps and determined distances for selected sources using dedicated observations of the associated $^{12}$CO(2--1) emission.
\subsection{Source extraction and selection}
\label{sec:sourceextraction}
The region in the outer Galaxy chosen for this survey ($225\degree \leq \ell \leq 260\degree$, $-3\degree \leq b \leq 0.5\degree$) was selected, as this part of the outer Galaxy has never been observed with high spatial resolution or sensitivity before. Furthermore, the sections of the spiral arms in this region are well separated in velocity, making it relatively straightforward to associate objects with their parent spiral arm. The latitude range was selected as the minimum and maximum latitudes covered by the Herschel Hi-GAL dust continuum emission maps to ensure availability of complementary data.
The Herschel SPIRE\,250\,\mum emission data obtained by Hi-GAL \citep{Molinari2010} was used to identify clumps in this region. Note that due to the Galactic warp the latitude range covered by Hi-GAL is centred around $b=-1\degree$ spanning \simm2.5\degree. Although the authors are aware of the Hi-GAL compact source catalogues that are publicly available for the inner Galaxy \citep{Molinari2016}, their outer Galaxy counterpart is not released yet. Therefore an independent method was chosen to obtain a source catalogue for the targeted area of the outer Galaxy. To identify emission peaks and obtain their source sizes, SExtractor \citep{Bertin1996} was used, as it was used by team members in a similar way to produce the ATLASGAL compact source catalogue \citep{Contreras2013,urquhart2014} with great success, and will allow for more reliable comparisons between the outer Galaxy and ATLASGAL datasets.
\begin{figure*}[tp!]
   \centering
   \includegraphics[width=0.95\linewidth]{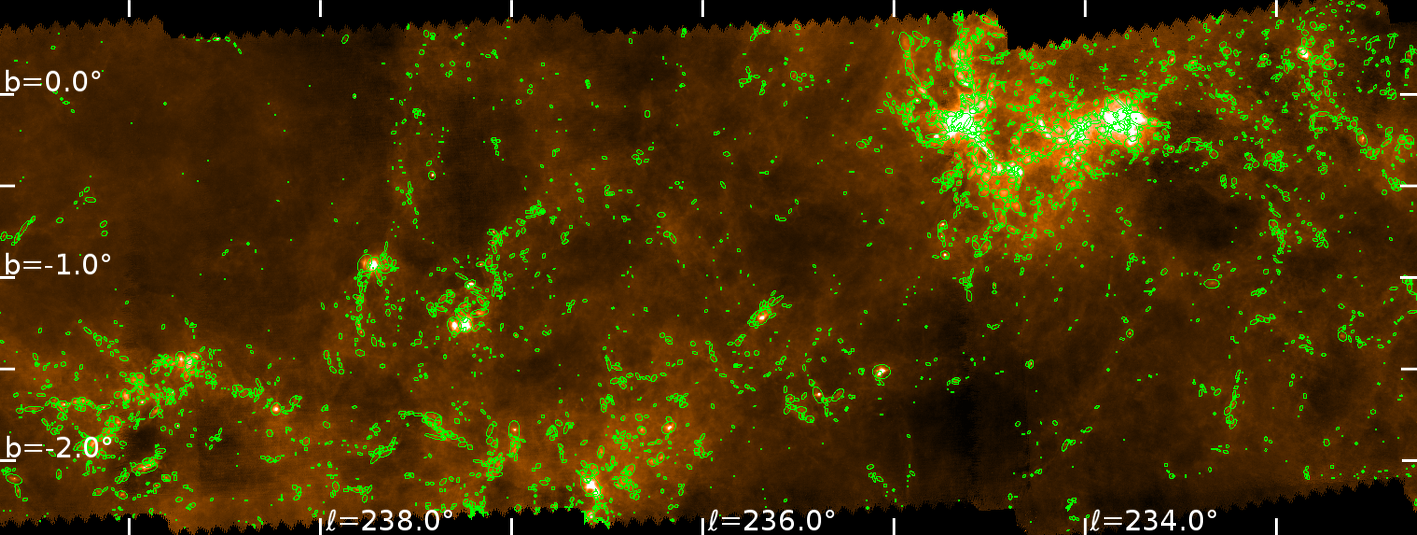}
      \caption[SPIRE\,250\,\micron emission showing an example of extracted sources]{Image of SPIRE\,250\,\micron emission showing an example of extracted sources around $\ell=236\degree$ and $b=-1\degree$. The image spans \simm7\degree\ in longitude and \simm3\degree\ in latitude. Over-plotted are the sources identified by SExtractor (green ellipses).}
         \label{fig:source_identification}
\end{figure*}
\begin{table}
\begin{center}
\small
\caption[Astrometric data and integrated flux for the extracted clumps]{Astrometric data and integrated flux as determined by SExtractor for the extracted clumps for the first 20 clumps from a total of 23,817 sources. Columns are as follows: name of the source, Galactic longitude, Galactic latitude, full-width-at-half-maximum source size and the 250\,\micron integrated flux as measured by SExtractor.} \begin{tabular}{cccccccc}
\hline
Name & $\ell$ & $b$ & FWHM & $F_{250}$ \\
& {[deg]} & {[deg]} & {[arcsec]} & {[Jy]} \\
\hline\hline
G225.002-00.589 & 225.002 & $-$0.589 & 14.4 & 3.65 \\
G225.003-00.574 & 225.003 & $-$0.574 & 21.2 & 7.78 \\
G225.003-01.113 & 225.003 & $-$1.113 & 14.4 & 3.85 \\
G225.004-00.245 & 225.004 & $-$0.245 & 10.1 & 2.69 \\
G225.004-00.280 & 225.004 & $-$0.280 & 12.2 & 4.43 \\
G225.004-00.281 & 225.004 & $-$0.281 & 14.4 & 2.49 \\
G225.004-01.150 & 225.004 & $-$1.150 & 8.6 & 5.48 \\
G225.013-00.319 & 225.013 & $-$0.319 & 15.1 & 6.79 \\
G225.013-00.570 & 225.013 & $-$0.570 & 8.6 & 6.39 \\
G225.014-00.570 & 225.014 & $-$0.570 & 14.4 & 7.03 \\
G225.014-01.125 & 225.014 & $-$1.125 & 14.4 & 3.65 \\
G225.014-01.126 & 225.014 & $-$1.126 & 24.5 & 5.33 \\
G225.015-01.682 & 225.015 & $-$1.682 & 8.3 & 2.56 \\
G225.015-01.683 & 225.015 & $-$1.683 & 10.8 & 2.64 \\
G225.016-00.586 & 225.016 & $-$0.586 & 18.0 & 5.11 \\
G225.016-01.588 & 225.016 & $-$1.588 & 25.2 & 3.53 \\
G225.017-00.587 & 225.017 & $-$0.587 & 22.3 & 5.03 \\
G225.017-01.588 & 225.017 & $-$1.588 & 15.8 & 3.42 \\
G225.021-00.301 & 225.021 & $-$0.301 & 26.3 & 7.77 \\
G225.022-00.301 & 225.022 & $-$0.301 & 18.0 & 6.46 \\
G225.024-00.178 & 225.024 & $-$0.178 & 18.0 & 2.22 \\
G225.027-00.982 & 225.027 & $-$0.982 & 7.2 & 2.32 \\

\hline
\end{tabular}
\label{tab:outer_co:sources_extracted}
\end{center}
\end{table}
As a general approach we searched for emission peaks that have at least 4 pixels (i.e. \simm1 beam) above a threshold of $3\sigma_\mathrm{rms}$ above the local background noise level. We used two different background mesh sizes of 64 or 32 pixels to either exclude or include the 6 brightest and most complex regions of emission, respectively, as these crowded regions would introduce a bias in the source selection process. As a result, we obtained two catalogues: one excluding the brightest regions, which includes 12,783 sources, and a second one that was optimized to identify clumps located in the bright regions, that contains 15,874 sources. Merging both catalogues and removing duplicate entries, we obtained a full catalogue identifying a total of 23,817 sources in the $225\degree \leq \ell \leq 260\degree$ region. The result of the source extraction  from the SPIRE\,250\,\mum\ images can be seen in Fig.\,\ref{fig:source_identification} which includes a variety of bright complexes and dark regions devoid of emission. In Table\,\ref{tab:outer_co:sources_extracted} we give the parameters of the extracted sources for a small portion of the catalogue.
\begin{figure*}[tp!]
   \centering
   \includegraphics[trim=0cm 0.4cm 0cm 0cm,clip,width=0.38\linewidth]{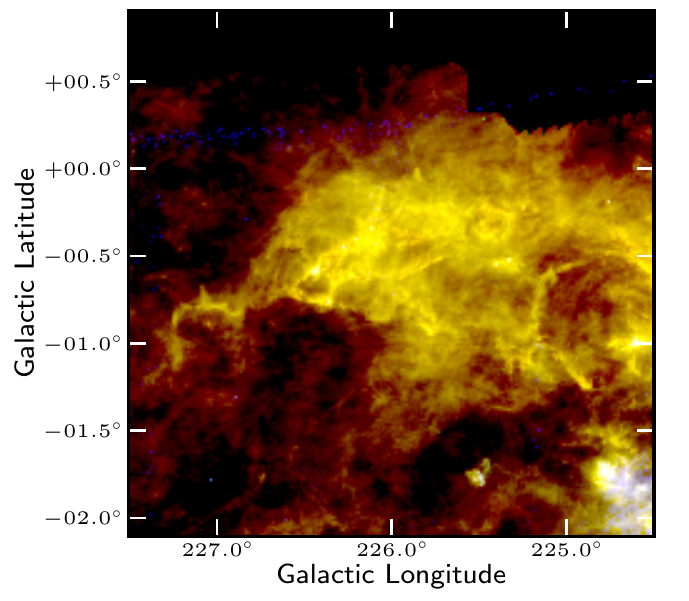}
   \includegraphics[trim=0cm 0.4cm 0cm 0cm,clip,width=0.38\linewidth]{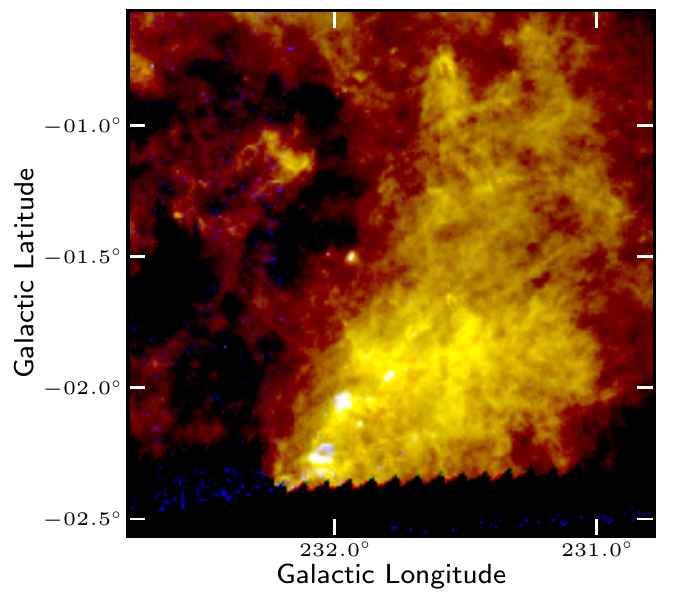}\\
   \includegraphics[trim=0cm 0.4cm 0cm 0cm,clip,width=0.38\linewidth]{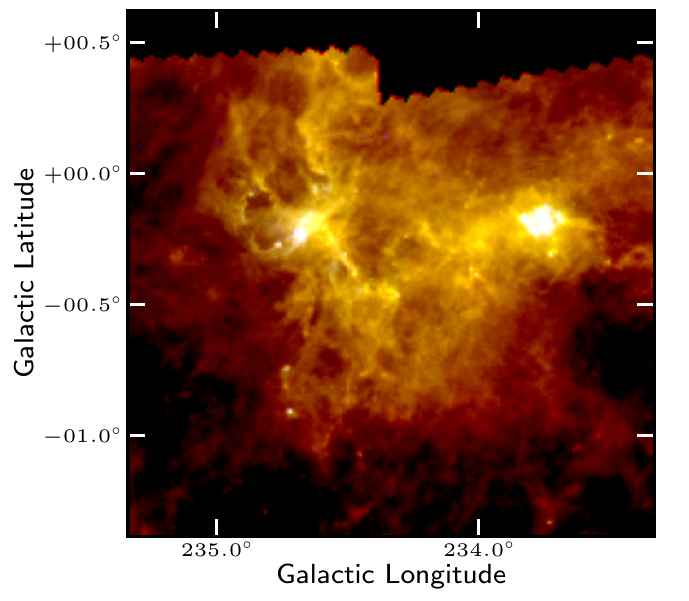}
   \includegraphics[trim=0cm 0.4cm 0cm 0cm,clip,width=0.38\linewidth]{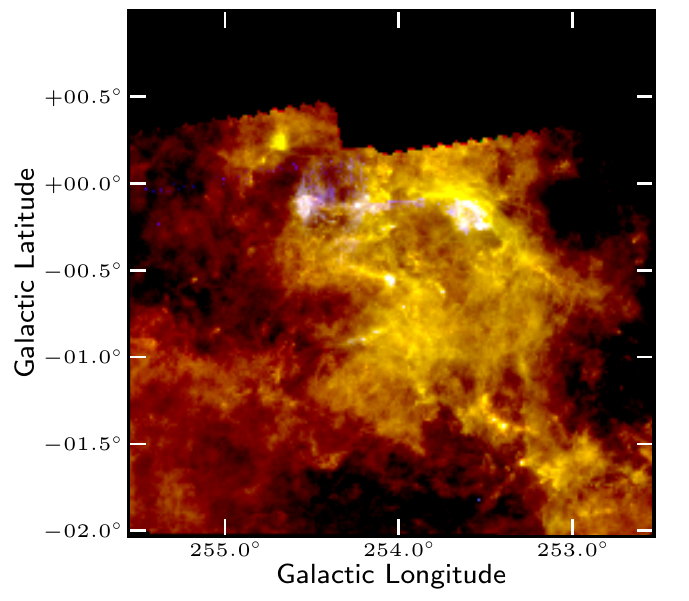}\\
   \includegraphics[width=0.38\linewidth]{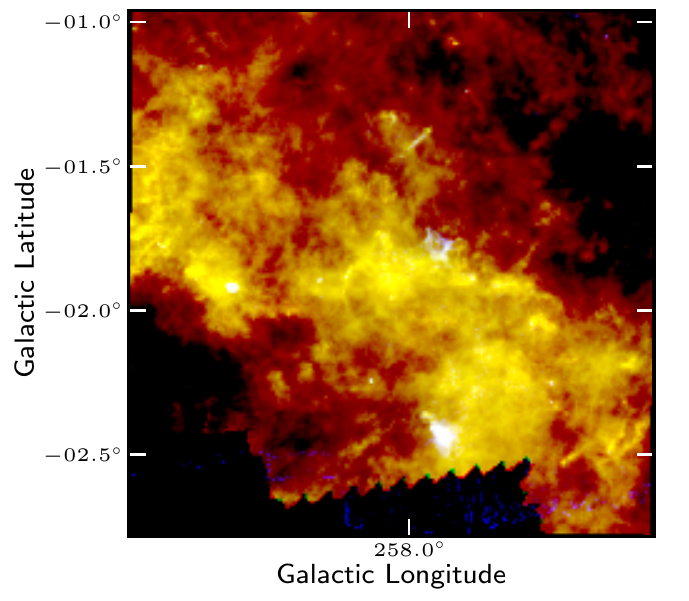}
   \includegraphics[width=0.38\linewidth]{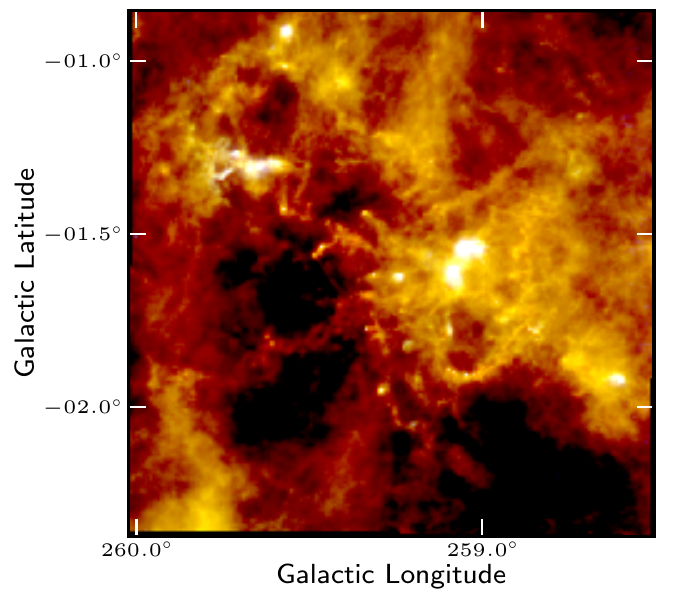}\\
      \caption{RGB images of the six brightest regions in the survey area. These have been excluded from the automated source selection, to avoid biasing the survey towards these regions. Red: SPIRE\,500\,\mum; Green: SPIRE\,250\,\mum; Blue: PACS\,70\,\mum.}
         \label{fig:outer:bright}
\end{figure*}
As one major goal of this research is to probe the full distance range from the nearest to the farthest sources, we used the subset excluding the 6 brightest regions to select the majority of our sources. However, in order to characterize these regions, too, we manually added 34 emission peaks from the full catalogue, ensuring these regions are included but not over-represented (see Table\,\ref{tab:complexes_excluded} and Fig.\,\ref{fig:outer:bright} for an overview of these 6 regions). For the observations outside the six brightest and most complex regions, we selected the 100 sources yielding the highest peak flux as well as the 100 sources with the highest integrated flux, accounting for 196 sources with 4 sources satisfying both criterias. An additional 587 sources were finally selected randomly from the brightness limited subset, picking up faint sources that are more likely to be located at farther heliocentric distances. In total, \varTotalSources\ sources of the extracted emission peaks were observed.
\begin{table}
\begin{center}
\small
\setlength{\tabcolsep}{5pt}
\caption{Overview of the 6 dominating bright and complex regions and the number of sources taken into account in the present paper.}
\begin{tabular}{@{}ccccccc@{}}
\hline
$\ell$ & $b$ & Count & $\varv_\mathrm{lsr}$ & $R_\mathrm{hel}$ & $R_\mathrm{gal}$ & Extent \\
{[deg]} & {[deg]} &   & {[km\,s$^{-1}$]} & {[kpc]} & {[kpc]} & {[deg]} ({[pc]})\\
\hline\hline
259.06 & $-$1.61 & 7 & 59.3 & 6.5 & 11.5 & 0.5 (56) \\
258.06 & $-$1.87 & 3 & 44.0 & 4.9 & 10.5 & 0.7 (59) \\
254.04 & $-$0.22 & 6 & 36.1 & 3.9 & 10.1 & 1.5 (100) \\
234.33 & $-$0.28 & 7 & 43.0 & 3.8 & 11.0 & 1.4 (92) \\
231.78 & $-$1.96 & 8 & 47.3 & 4.2 & 11.4 & 0.9 (66)\\
225.50 & $-$0.40 & 8 & 16.0 & 1.4 & 9.4 & 1.1 (25)\\
\hline
\end{tabular}
\label{tab:complexes_excluded}
\setlength{\tabcolsep}{5pt}
\end{center}
\end{table}
\begin{figure}[tp!]
   \centering
   \includegraphics[scale=0.9]{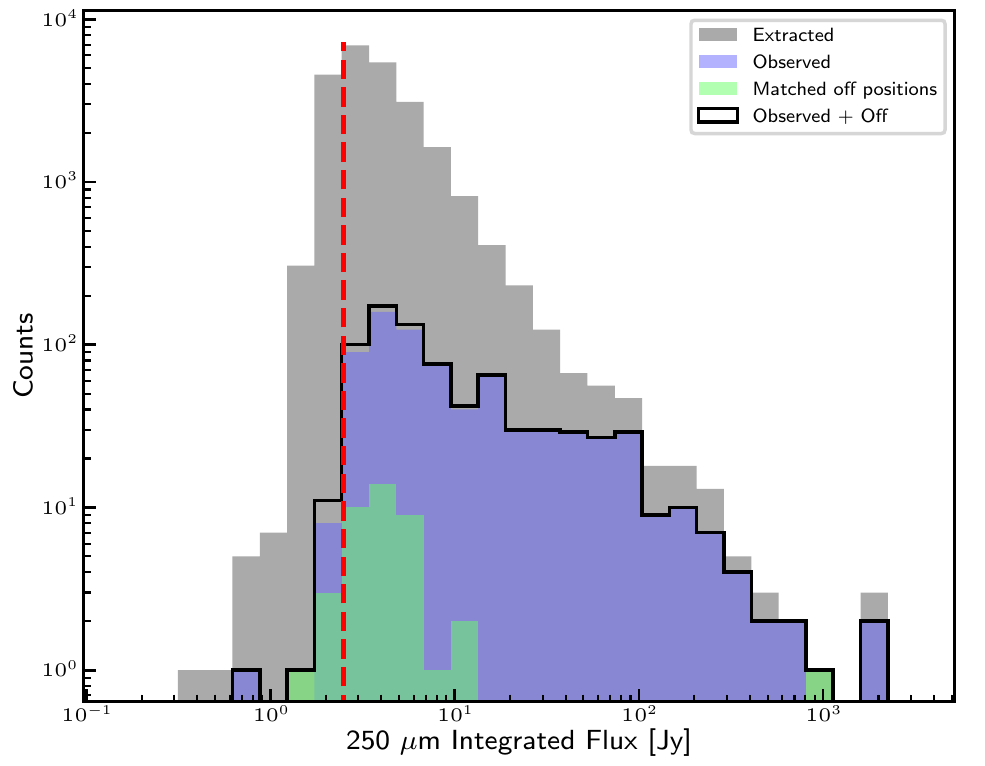}
      \caption{Histogram showing the integrated flux for the sources identified by SExtractor (grey) in comparison to those selected for our observations (blue) and those recovered from contaminated off-source positions (green; see Sect.\,\ref{sect:offpos}). The red vertical dashed line indicates the limit of \simm2.5\,Jy, that was imposed as a selection threshold (see text for details).}
         \label{fig:source_selection}
\end{figure}
To show that our source selection procedure does not introduce a strong flux bias, we show in Fig.\,\ref{fig:source_selection} the histogram of the aperture flux at 250\,\mum of all extracted sources (including the brightest regions) together with the histogram of the aperture flux of the observed sources. As can be seen, our observed sources sample the whole range of the flux distribution of the complete catalogue very well down to the limit of $\sim$2.5\,Jy, which was imposed as a selection threshold to ensure the $^{12}$CO(2--1) line could be detected at a $3\sigma$ level in a reasonable amount of integration time ($\lesssim3$\,min).
\subsection{Observations}
\label{sect:outer:co_obs}
Carbon monoxide (CO) is the second most abundant molecule in the Milky Way after H$_2$. It is widely found in molecular clouds, and even more so in the denser parts constituting the dusty clumps we are investigating in this work. Observing the CO emission towards a dust clump allows us to obtain a velocity for a given clump and from this infer a distance to the clump.
We used the facility receiver APEX-1 \citep[SHeFI; ][]{vassilev2008} at the Atacama Pathfinder Experiment 12\,m submillimeter telescope \citep[APEX; ][]{gusten2006} to obtain line-of-sight velocities ($\varv_\mathrm{lsr}$) of the selected sources. Two wide-band Fast Fourier Transform Spectrometers \citep[FFTS; ][]{Klein2012} make up the back-ends, each consisting of 32,768 spectral channels covering an instantaneous bandwidth of 2.5\,GHz. We targeted the $^{12}$CO(2--1) transition at 230.5\,GHz, as line-of-sight confusion is not as much an issue for the outer Galaxy than for the inner Galaxy, and the higher column densities allow for detection of sources at greater distances with similar integration times when compared to the rarer and less bright $^{13}$CO(2--1) and C$^{18}$O(2--1) transitions. With this setup we were able to observe the targeted transition at a velocity resolution of $\sim0.1$\,km\,s$^{-1}$.
\begin{table}[tp!]
\begin{center}
\small
\caption[Summary of the APEX observational parameters for the outer Galaxy]{Summary of the APEX observational parameters.}\label{tab:obs_parameters}
\begin{tabular}{lc}
\hline
Parameter & Value \\
\hline\hline
Galactic longitude range      &      225\degr\ $< \ell <$  260\degr\ \\
Galactic latitude range      &      $-$2.73\degr\  $< b <$ 0.45\degr\ \\
Number of observations & \varTotalSources \\
Number of spectra with emission & \varSourcesWithEmission \\
Instrument & SHeFI (APEX-1) \\
Frequency & 230.538\,GHz \\
Angular resolution & 30\arcsec \\
Spectral resolution & 0.1\,\kms\ \\
Smoothed spectral resolution & 1\,\kms\ \\
Mean noise (T$^{*}_{\rm{A}}$) & $\sim$50\,mK\,channel$^{-1}$ \\
Average PWV  & $\sim$2.7\,mm\\
Integration time (on-source) & \varIntegrationTime\,mins\\
\hline
\end{tabular}
\end{center}
\end{table}
The observations towards \varTotalSources\ sources were conducted between July 2013 and December 2016 as a bad weather backup project at APEX with precipitable water vapour (PWV) up to 6\,mm. To reach an $\mathrm{rms}<0.1$~K per channel, each source was observed in position switching mode for 3--6~min, depending on PWV, elevation and line-strength. The off-positions were selected as relative offsets with a separation of 1\degree\ roughly perpendicular to the Galactic plane. The pointed $^{12}$CO(2--1) observations were obtained with an average PWV of 2.7\,mm, resulting in an average RMS of 0.05\,K at a smoothed channel resolution of 1\,km\,s$^{-1}$. Sample spectra are shown in Fig.\,\ref{fig:co_plots}. In total, we found \varSourcesNoEmission\ spectra without any emission, leaving \varSourcesWithEmission\ (\simm\varSourcesWithEmissionFraction\%) spectra for further analysis. A summary of all observational parameters is given in Table\,\ref{tab:obs_parameters}. To obtain CO(2--1) velocity information for each clump, we applied the methods described in the following section.
\subsection{Data reduction procedures}
\subsubsection{Identifying velocity components}
The obtained spectra were reduced using the Continuum and Line Analysis Single-dish Software (CLASS\footnote{GILDAS/CLASS: \url{https://www.iram.fr/IRAMFR/GILDAS/}}). To obtain the velocity components from the observed spectra we first combined all scans for a single observed position into a single spectrum. This was successively smoothed to a velocity resolution of 1\,\kms\ and a linear baseline was subtracted. The spectra were then subjected to a Python code, where they were limited to the velocity range of $-$20\,\kms\  < $\varv_\mathrm{lsr}$ < 150\,\kms, in order to limit the spectra to velocity range expected for the outer Galaxy. Coherent groups of emission and absorption likely associated with a single cloud were determined by defining a window where all emission is above the $3\sigma$ noise level. These emission groups are then fitted iteratively, starting with the brightest group. First the spectrum is de-spiked, after which the number of peaks in a group is determined and Gaussian profiles are fitted to the peaks, and the resulting fit is subtracted from the original spectrum. The procedure is then repeated until no residual emission above $3\sigma$ is found and all groups are fitted. In order to avoid adding too many emission components that are close to each other and are likely associated with the same cloud, we only considered peaks separated by at least twice the width of the fitted Gaussian as major emission components. The same process was repeated for negative emission features, allowing us to identify observations with a contaminated off-position.
\begin{figure*}[tp!]
   \centering
   \includegraphics[trim=-0.2cm 0.0cm 0cm 0cm,clip,width=0.49\linewidth]{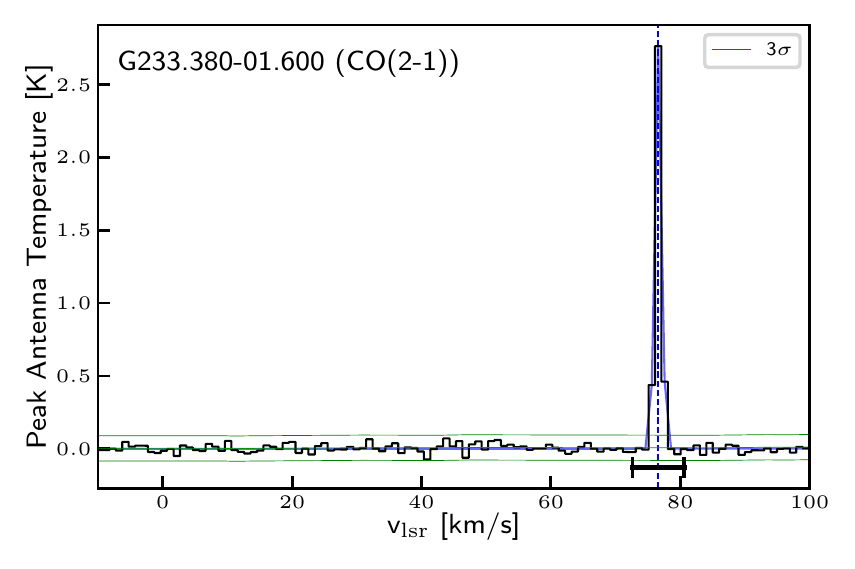}
   \includegraphics[width=0.49\linewidth]{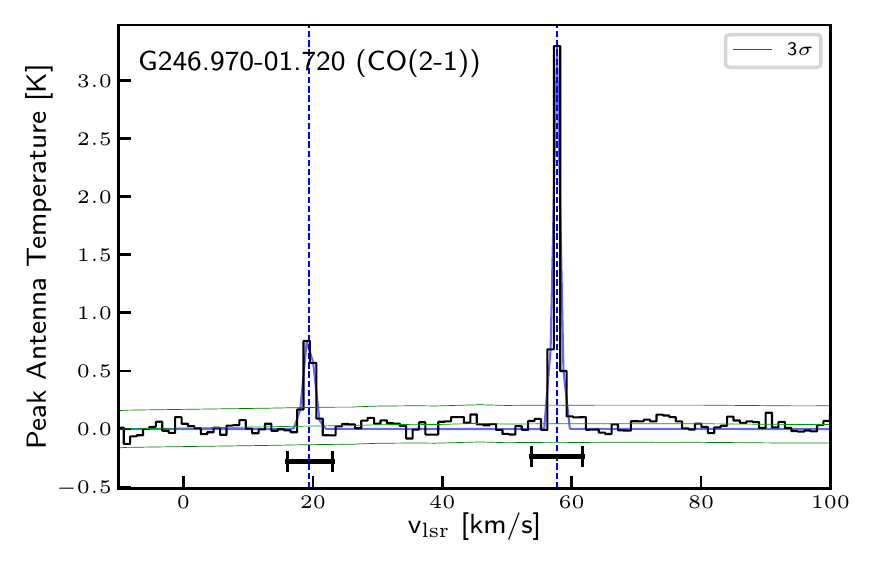}\\
   \includegraphics[width=0.49\linewidth]{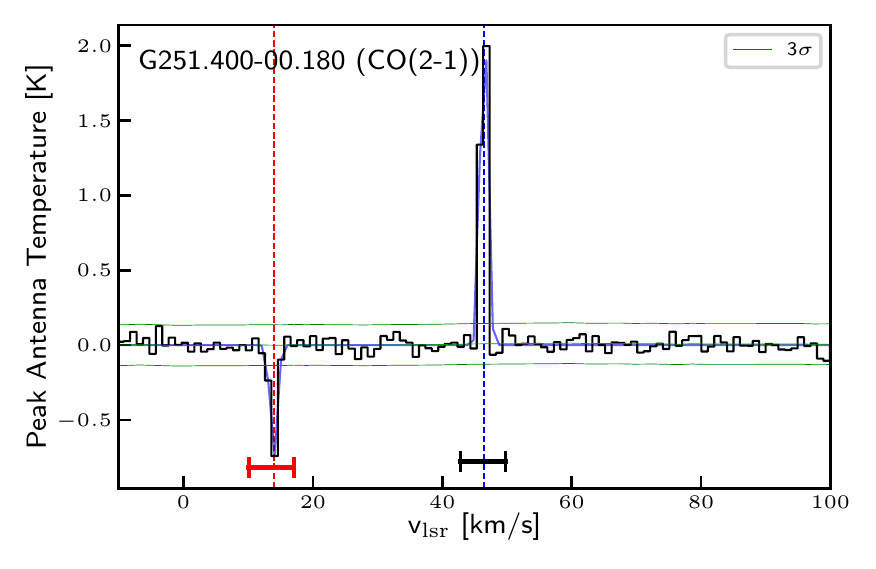}
   \includegraphics[width=0.49\linewidth]{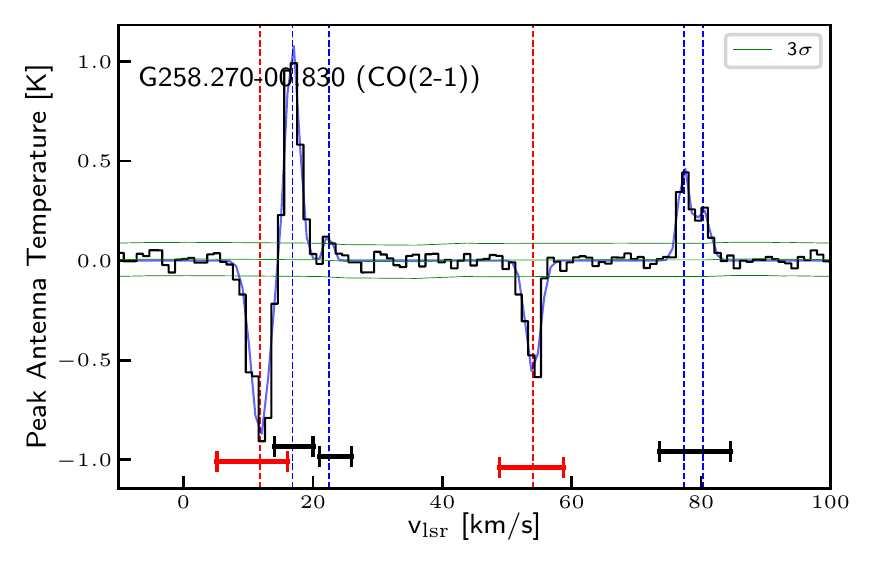}
      \caption[Four $^{12}$CO(2--1) spectra showing different typical profiles]{$^{12}$CO(2--1) spectra (black line) showing different typical profiles: single component (upper left), multiple components (upper right), simple contamination in the off-position (lower left) and a complex spectrum with all features (lower right). The total fitted spectrum is overplotted in blue. Coherent velocity complexes are either marked as black (source position) or red (off position; negative features) horizontal bars below the spectrum. The vertical dashed blue lines indicate the peak positions of the fitted Gaussian profiles, whereas the vertical red dashed lines indicates the peak position of the emission found in the off-position.}
         \label{fig:co_plots}
\end{figure*}
In Fig.\,\ref{fig:co_plots} we show example spectra of velocity measurements obtained towards sources located in the outer Galaxy. In the upper left panel a single emission component along the line-of-sight can be fitted with a single Gaussian profile. This allows the velocity to be immediately assigned to the dust clump without further analysis. The situation becomes more difficult when a cloud is either composed of multiple components or there are multiple clouds located along the line of sight (Fig.\,\ref{fig:co_plots}, upper right panel). Furthermore, as CO is the second most abundant molecule in the Milky Way, the reference position used when taking the spectrum might be contaminated, resulting in negative features in the spectrum (Fig.\,\ref{fig:co_plots}, lower left panel). Contaminated spectra pose the problem that the emission might be at a similar velocity as the negative feature, thus rendering the velocities less reliable. In extreme cases all these three effects are present in a single spectrum taken (Fig.\,\ref{fig:co_plots}, lower right panel).
Although these effects complicate the analysis, in most cases still a velocity can be assigned to the corresponding clump. Assuming that the emission as seen in the continuum maps is associated mainly with the brightest emission found in CO, only the CO emission with the highest integrated intensity is taken into account. If this CO cloud has an integrated intensity at least twice as high as the rest of the CO emission, we assume that the dust emission is associated with it. In case a negative feature is located close to it ($\mathrm{d}v<1$\,\kms), the uncertainty of the velocity measurement is increased by the width of the negative feature from the off-position, considering the velocity measurement is still usable but with a higher uncertainty.
With the line-of-sight velocity of a clump known, we can determine its distance from the rotation curve of the Galaxy. As there is only one rotation direction in the southern outer Galaxy, namely from higher to lower longitudes, no distance ambiguity exists (in contrast to the inner Galaxy where this poses a problem), allowing for a direct conversion from the observed line-of-sight velocities to heliocentric distance.
We were able to fit spectra for \varSourcesWithEmission\ (\varSourcesWithEmissionFraction \%) of the observed \varTotalSources\ lines-of sight, identifying a total of \varNumComplexesOn\ clouds consisting of \varNumComponentsOn\ velocity components from the CO(2--1) emission above the 5$\sigma$ level. From here on we refer to a single fitted Gaussian as a velocity component and to any coherent group of emission as a cloud, as we assume that these groups of emission are physically connected with similar velocities but distinct components.
We found the source-off reference positions, selected as a relative offset from the target position of one degree, to be contaminated for \varNumPosOff\ sources of the sample (\varNumPosContaminatedFraction\%), as shown for two lines-of-sight in the lower panels of Fig.\,\ref{fig:co_plots}. As the uncertainty in distances determined from the $\varv_\mathrm{lsr}$ are mostly dominated by the uncertainty in the rotation curve used for the calculation rather than from the velocity, we assume that an uncertainty in $\varv_\mathrm{lsr}$ of a few \kms due to contamination in the source-off position, does not pose a serious problem. In fact, we conclude that the CO emission at the source-off position can be used to increase the number of positions for velocity measurements towards the outer Galaxy, but care has to be taken on their usage as discussed in the following paragraph.
\label{sect:offpos}
Therefore we also analyse the emission at the off positions when a source shows contamination. To do so, we simply invert the baseline corrected spectra and apply the same procedure as for the emission of uncontaminated sources, yielding velocity components for the off-position. We want to stress that these additional CO(2--1) clouds do not necessarily coincide with the peak of any dust clump or cloud, but as these components are clearly above the background noise level, they improve our view on the large-scale structures in the outer Galaxy. In contrast, for the SED analysis of the dust clumps in Section\,\ref{sect:outer:seds}, we only take into account off-source positions that are matched with a SExtractor source position within one beam size (i.e. 30\arcsec), yielding an additional 40 sources for which we can derive physical properties. If an absorption feature was found within 10\,\kms\ of an emission peak, the uncertainties for all derived properties were increased by a factor of two (i.e. velocity, peak temperature and line-width). We choose a rather large range of 10\,\kms, in order to make sure that we also cover cases with broad emission complexes, where the emission between two peaks cancels each other out. 
Analysing the emission at the off positions for the \varNumPosOff\ contaminated spectra, we were able to add another \varNumComplexesOff\ clouds above the $5\sigma$ noise level from \varNumLosOffComplex\ lines-of-sight to our analysis. This increases the total number of clouds by \varNumComplexesOffFraction \% from \varNumComplexesOn\ to \varNumComplexes\ consisting of a total of \varNumComponents\ components and covering a total of \varNumPos\ lines-of-sight (targeted + reference). We present the results for each position in Table\,\ref{tab:velocities}.
As observations were often conducted under bad-weather conditions with high and fast varying PWV sometimes in excess of 5\,mm, we calculate the main-beam brightness temperature only for sources that have a calibration within a certain time-frame according to the weather conditions\footnote{within 15, 10, 7 or 5 minutes, for PWV$<1.5$, $<3$, $<4$, $\geq4$\,mm, assuming increasing weather stability with lower PWV.}, resulting in the removal of  \varNumBadCal\ lines-of sight (\varNumBadCalComplexes\ clouds). For the remaining sources the main beam temperature $T_\mathrm{mb}$ was then calculated from the antenna temperature $T^*_\mathrm{A}$ multiplying with a forward efficiency of $\eta_\mathrm{f}=95$\% and dividing by a beam efficiency of $\eta_\mathrm{mb}=75$\%. In case no calibration was found within the given time-frame, we do not give a main beam temperature or intensity.
For \varNumLosSingleComplexes\ (\varNumLosSingleComplexesFraction\%) lines-of-sight only a single cloud is found, whereas multiple clouds are identified for \varNumLosMultiComplexes\ (\varNumLosMultiComplexesFraction\%) lines-of-sight. For \varSourcesNoEmission\ lines-of-sight no emission was found. In summary we identified a total of \varNumGoodCalComplexes\ clouds for all \varNumPos\ lines-of-sight, yielding an average of \varNumComplexesPerLoS\ clouds per line-of-sight.
We give all clouds (i.e.\ groups of emission) in Table\,\ref{tab:vel_complexes}, which we define as continuous emission features above $3\sigma$\,rms, consisting of one or more fitted Gaussians. An example of two components (possibly arising from self absorption) making up one cloud can be seen in Fig.\,\ref{fig:co_plots} (lower right) at \simm80\,\kms indicated by the solid black bar at the lower part of the figure. The velocity found for the brightest velocity complex along a given line-of-sight is later used to assign a distance to the clump as seen in continuum emission, from which the physical parameters are derived.
\setlength{\tabcolsep}{5pt}
\begin{table*}
\begin{center}
\small
\caption[Velocity components identified from the CO(2--1) observations]{Velocity components identified from the CO(2--1) observations along 10 lines-of-sight. Source names starting with a `\textit{G}' are measured emission at targeted positions, whereas names starting with an `\textit{O}' indicate emission in the off-position. Uncertainties for $\varv_\mathrm{lsr}$ are in the order of 0.5\,\kms, distance uncertainties in the order of 0.3\,kpc.}
\label{tab:velocities}
\begin{tabular}{
@{}
c
c
S[table-format=1.1]
S[table-format=1.1]
r
r
S[table-format=1.1]
S[table-format=1.1]
@{}
}
\hline
{Line-of-sight} & Cloud-Component & {$\varv_\mathrm{lsr}$} & {$\Delta v_\mathrm{lsr}$} & \multicolumn{1}{c}{$T^*_\mathrm{A, max}$} & \multicolumn{1}{c}{$\int{T dv}$} & {$R_\mathrm{hel}$} & {$R_\mathrm{gal}$}\\
 & & {(km s$^{-1}$)}& {(km s$^{-1}$)} & \multicolumn{1}{c}{(K)} & \multicolumn{1}{c}{(K km s$^{-1}$)} & {(kpc)} & {(kpc)}\\
\hline
G232.420+00.240 & 1-1 & 20.6 & 1.0 & $1.9 \pm 0.1$ & $0.2 \pm 4.7$ & 1.8 & 9.5 \\
G232.490$-$00.300 & 1-1 & 17.0 & 0.9 & $2.1 \pm 0.1$ & $0.3 \pm 4.5$ & 1.5 & 9.3 \\
G232.500$-$00.040 & 1-1 & 20.2 & 0.7 & $0.9 \pm 0.2$ & $0.3 \pm 1.6$ & 1.8 & 9.5 \\
  & 1-2 & 17.3 & 0.9 & $2.2 \pm 0.1$ & $0.2 \pm 5.3$ & 1.5 & 9.3 \\
G232.504$-$01.223 & 1-1 & 25.3 & 0.6 & $0.4 \pm 0.1$ & $0.1 \pm 0.6$ & 2.2 & 9.8 \\
  & 2-1 & 47.3 & 0.6 & $0.3 \pm 0.1$ & $0.1 \pm 0.4$ & 4.2 & 11.4 \\
G232.510+00.200 & 1-1 & 20.9 & 1.1 & $1.6 \pm 0.2$ & $0.2 \pm 4.2$ & 1.8 & 9.6 \\
  & 1-2 & 15.9 & 0.7 & $1.6 \pm 0.2$ & $0.4 \pm 2.8$ & 1.4 & 9.3 \\
  & 1-3 & 13.8 & 0.3 & $0.8 \pm 0.2$ & $0.5 \pm 0.6$ & 1.3 & 9.2 \\
G232.517$-$02.947 & 1-1 & 34.1 & 0.6 & $0.2 \pm 0.1$ & $0.1 \pm 0.3$ & 3.0 & 10.4 \\
G232.590$-$01.190 & 1-1 & 22.5 & 0.6 & $0.7 \pm 0.1$ & $0.1 \pm 1.1$ & 1.9 & 9.6 \\
G232.596$-$01.352 & 1-1 & 48.4 & 0.7 & $2.3 \pm 0.3$ & $1.0 \pm 4.1$ & 4.3 & 11.5 \\
  & 1-2 & 46.0 & 1.0 & $3.0 \pm 0.2$ & $0.6 \pm 7.6$ & 4.1 & 11.3 \\
G232.600$-$01.320 & 1-1 & 47.1 & 1.5 & $2.0 \pm 0.1$ & $0.1 \pm 7.4$ & 4.2 & 11.4 \\
G232.600+00.430 & 1-1 & 16.3 & 0.6 & $3.9 \pm 0.1$ & $0.7 \pm 5.7$ & 1.5 & 9.3 \\
G232.600+00.100 & 1-1 & 17.0 & 1.7 & $1.9 \pm 0.1$ & $0.1 \pm 8.4$ & 1.5 & 9.3 \\
G232.600+00.300 & 1-1 & 15.7 & 0.6 & $4.7 \pm 0.1$ & $1.1 \pm 7.3$ & 1.4 & 9.3 \\
G232.630+00.390 & 1-1 & 16.6 & 0.6 & $3.3 \pm 0.1$ & $0.8 \pm 4.7$ & 1.5 & 9.3 \\
G232.640$-$00.420 & 1-1 & 23.2 & 0.5 & $2.7 \pm 0.2$ & $1.0 \pm 3.7$ & 2.0 & 9.7 \\
  & 2-1 & 47.4 & 0.6 & $1.0 \pm 0.1$ & $0.2 \pm 1.3$ & 4.2 & 11.4 \\
  & 3-1 & 17.7 & 1.4 & $0.4 \pm 0.1$ & $0.0 \pm 1.2$ & 1.6 & 9.4 \\

\hline
\end{tabular}
\end{center}
\end{table*}
\setlength{\tabcolsep}{6pt}
\begin{table*}
\begin{center}
\small
\caption{Clouds (i.e. coherent groups of velocity components) identified from the CO(2--1) observations along a given line-of-sight for 15 complexes. Source names starting with a `\textit{G}' are measured emission at targeted positions, whereas names starting with an `\textit{O}' indicate emission in the off-position.}
\label{tab:vel_complexes}
\begin{tabular}{
@{}
c
c
c
S[table-format=1.1]
S[table-format=1.1]
r
r
@{}
}
\hline
{Line-of-sight} & {Cloud} & {\# Components} & {$\varv_\mathrm{lsr}$} & {$\Delta v_\mathrm{lsr}$} & \multicolumn{1}{c}{$T^*_\mathrm{A, max}$} & \multicolumn{1}{c}{$\int{T dv}$} \\
 & & & {(km s$^{-1}$)} & {(km s$^{-1}$)} & \multicolumn{1}{c}{(K)} & \multicolumn{1}{c}{(K km s$^{-1}$)} \\
\hline
G232.057$-$01.163 & 1/1 & 1 & 45.7 & 5.0 & $0.2 \pm 0.1$ & $0.5 \pm 0.3$ \\
G232.062$-$00.126 & 1/3 & 1 & 19.2 & 2.0 & $1.0 \pm 0.2$ & $1.2 \pm 0.2$ \\
G232.062$-$00.096 & 1/2 & 1 & 57.0 & 6.0 & $0.9 \pm 0.1$ & $1.3 \pm 0.1$ \\
  & 2/2 & 1 & 24.7 & 5.0 & $0.3 \pm 0.1$ & $0.6 \pm 0.2$ \\
G232.062$-$00.076 & 2/3 & 1 & 56.4 & 5.0 & $1.1 \pm 0.1$ & $1.8 \pm 0.2$ \\
  & 3/3 & 1 & 19.6 & 3.0 & $0.2 \pm 0.1$ & $0.2 \pm 0.2$ \\
G232.071$-$00.176 & 1/2 & 1 & 19.7 & 10.9 & $1.1 \pm 0.8$ & $1.9 \pm 1.4$ \\
G232.075$-$02.276 & 1/1 & 1 & 41.8 & 32.7 & $22.0 \pm 0.2$ & $116.1 \pm 1.8$ \\
G232.080+00.130 & 1/1 & 1 & 57.2 & 7.9 & $2.4 \pm 0.1$ & $5.4 \pm 0.2$ \\
G232.082$-$00.046 & 1/3 & 1 & 13.9 & 4.0 & $0.7 \pm 0.1$ & $0.9 \pm 0.1$ \\
  & 2/3 & 1 & 56.3 & 6.0 & $0.6 \pm 0.1$ & $1.1 \pm 0.2$ \\
G232.111$-$01.158 & 1/1 & 1 & 46.7 & 9.9 & $2.9 \pm 0.1$ & $12.4 \pm 0.5$ \\
G232.131$-$00.256 & 1/1 & 1 & 21.7 & 6.0 & $0.7 \pm 0.1$ & $1.1 \pm 0.2$ \\
G232.205$-$00.764 & 1/1 & 1 & 44.4 & 6.0 & $1.0 \pm 0.1$ & $1.3 \pm 0.1$ \\
G232.220+00.190 & 1/1 & 2 & 15.3 & 9.9 & $2.9 \pm 0.1$ & $6.6 \pm 0.3$ \\
G232.220$-$01.070 & 1/1 & 1 & 43.7 & 8.9 & $1.7 \pm 0.1$ & $7.1 \pm 0.4$ \\

\hline
\end{tabular}
\end{center}
\end{table*}
\subsection{Source velocities}
\begin{figure}[tp!]
   \centering
   \includegraphics[]{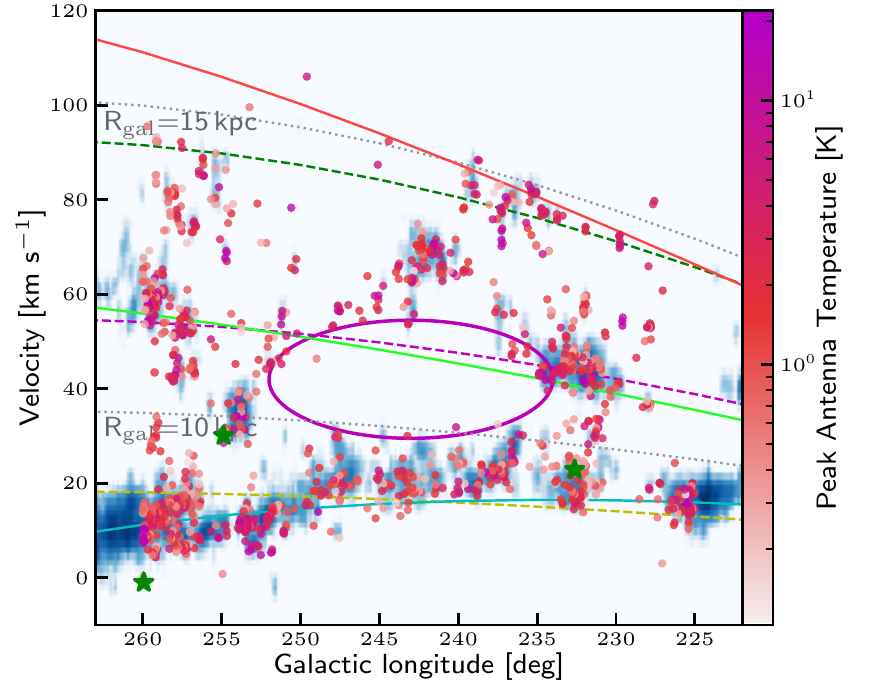}
      \caption[Galactic longitude vs. radial velocity]{Galactic longitude vs. radial velocity. The dots mark all clouds found above $5\sigma$ rms between $260\degree \leq \ell \leq 225\degree$, with the peak antenna temperature colour coded. Background image: CO(1--0) emission from \citet{Dame2001}. Dashed yellow, magenta, and green lines: peaks at 9, 11, and 14.0\,kpc, respectively, as found in Fig.\,\ref{fig:rgal_hist}. The spiral arms and the position of the local emission as determined from \hi emission are marked by the coloured solid lines. Cyan: local emission; green: Perseus arm; red: Outer arm. The solid magenta ellipse marks the rim of the Galactic supershell GSH\,242-3+77. Grey dashed lines: Galactocentric radii at 10 and 15\,kpc. The green stars mark the positions of the 3 sources from the Methanol Multibeam \citep[MMB;][]{green2012_mmb_outer} survey found in the region.}
         \label{fig:LV}
\end{figure}

In Fig.\,\ref{fig:LV} we present the longitude-velocity ($\ell$-$v$) plot covering the observed field. The velocity components above the $5\sigma$ rms noise level for all positions are overlaid on the $^{12}$CO(1--0) emission data cube from \citet{Dame2001} integrated for $-3\deg \leq b \leq 0.5\deg$ as approximately covered by our sample, that is following the Galactic warp. The loci of the spiral arms as well as the local emission are indicated by the solid line (see Sect.\,\ref{sect:spiral_arms} for details).
As can be seen, although we randomly sampled, components obtained for the CO(2--1) line trace all the major structures found in the CO(1--0) emission map very well. In addition, due to the 15 times higher resolution of our pointed observations when compared to the data from \citet[][i.e. 7.5\arcmin\ vs. 30\arcsec]{Dame2001}, we are able to trace small structures that have been previously missed, as these would fall below the detection limit due to beam dilution or blending with other clumps within the beam. For example, the structures with the highest velocity components at any given longitude would trace the most distant arm. The structure traced by our pointed observations from $\ell \approx 242 \degree$, \vlsr\,$ \approx 60$\,\kms\ to $\ell \approx 252 \degree$, \vlsr\,$ \approx 40$\,\kms\ is completely missed by the lower sensitivity CO(1--0) map. Additional features can be identified, indicating the structure between the spiral arms towards this region of the outer Galaxy to be more complex than indicated by the data from \citet{Dame2001}. We will discuss this in more detail in Section\,\ref{sect:structures}.
\begin{figure}[tp!]
   \centering
   \includegraphics[scale=0.9]{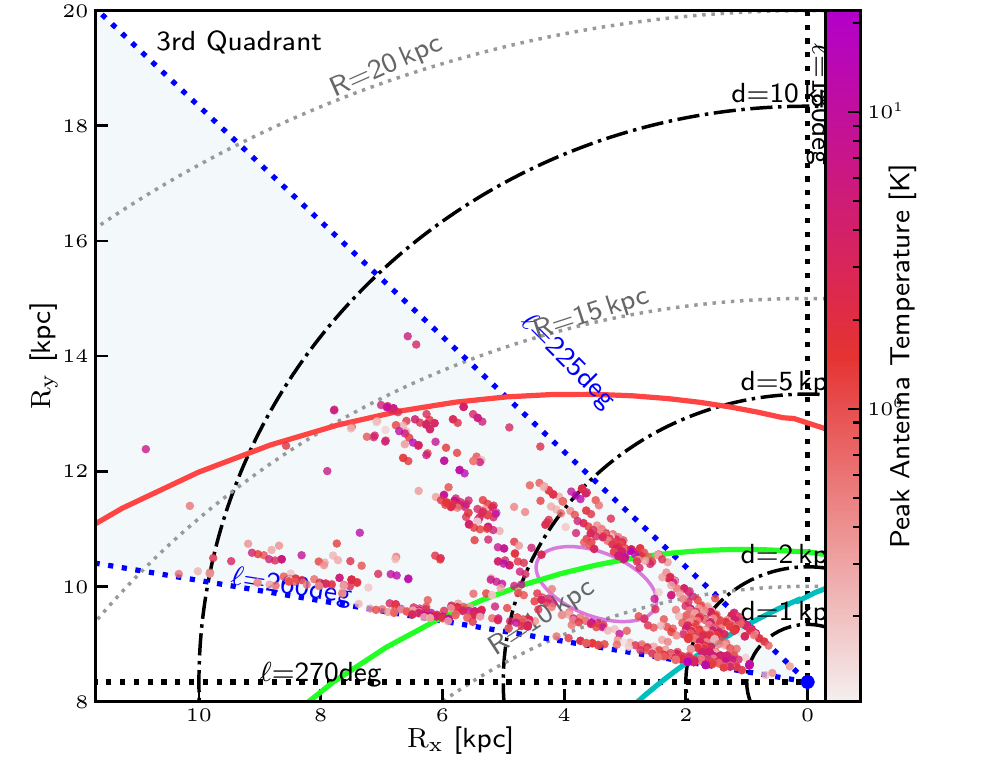}
      \caption[Third Quadrant of the Milky Way showing observed CO components]{Third Quadrant of the Milky Way. The dots mark all clouds above $5\sigma$ rms, using the same colour coding as Fig.\,\ref{fig:LV}. The observed region ($225\degree < \ell < 260\degree$) is highlighted in light blue. Heliocentric distances $d$ and Galactocentric radii $R$ are indicated as dash-dotted black and dotted grey circles, respectively.  Positions of arms and local emission are indicated by solid lines in the same colors as in Fig.\,\ref{fig:LV}. The position of the Galactic supershell GSH\,242-3+37 is indicated by the magenta ellipse. From the observed \vlsr\ we calculated the Heliocentric distances for each cloud using the \citet{Brand1993} rotation curve to obtain Galactic positions.}
	\label{fig:distr}
\end{figure}
\begin{figure}[tp!]
   \centering
   \includegraphics[]{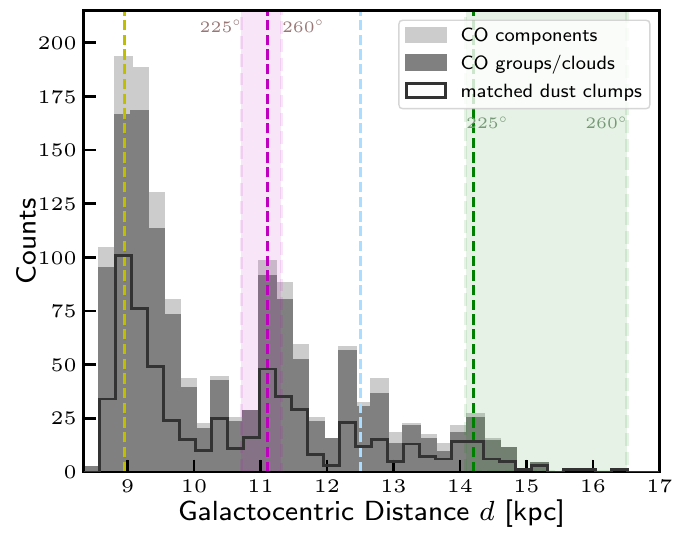}
   \includegraphics[]{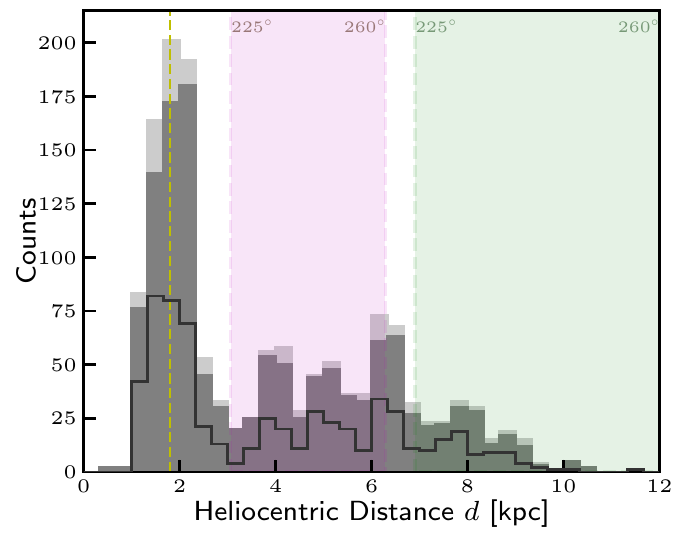}
      \caption[Histograms showing CO components, CO clouds and matched dust clumps]{Histograms showing CO velocity components above $5\sigma$ (light grey), CO emission groups/clouds (dark grey) and matched dust clumps (black outline) by Galactocentric (upper) and heliocentric (lower) distance. The vertical dashed yellow line indicates the peak associated with local emission. The magenta and green shaded areas mark the distance range between $225\degree \leq \ell \leq 260\degree$ of the loci of the Perseus and Outer arms, respectively. The two peaks between $r_\mathrm{gal}\sim12$ and $\sim13$\,kpc correspond to complexes between the Perseus and Outer arm.}
         \label{fig:rgal_hist}
\end{figure}
\subsection{Kinematic distances and Galactic distribution}
\label{sec:distances}
We determined distances for all velocity components by applying the \citet{Brand1993} rotation curve, assuming a distance to the Galactic centre of $R_0=8.34$\,kpc and an orbital velocity of $\theta_0=240$\,\kms, as derived by \citet{Reid2014} and used by \citet{Urquhart2018}. We prefer the rotation curve from \citet{Brand1993} over the more recent one from \citet{Reid2014}, as the latter is derived from parallactic distances to objects visible from the northern hemisphere and does not include measurements for the southern outer Galaxy.
In Fig.\,\ref{fig:distr} we show the distribution of our sources in the 3rd Galactic quadrant as viewed from the Galactic north pole. We find the closest source to be located at a heliocentric distance of $R_\mathrm{hel}$\,=\,\coRhelMin\,kpc and the most distant sources at $R_\mathrm{hel}$\,\simm\ 12\,kpc. This results in the sources spanning a range of Galactocentric distances between $R_\mathrm{gal}$\,=\,\coRgalMin\,kpc and $R_\mathrm{gal}$\,\simm\,16.5\,kpc. Note that the distances are not only affected by the uncertainty of the velocity measurement, but also by the spread due to streaming motions \citep[$\lesssim$10\,\kms;][]{Reid2014}, the expanding supershell \citep[$\sim$7\,\kms;][]{McClure-Griffiths2006}, and the accuracy of the rotation curve, which can change the distance of any given source as determined from the rotation curve significantly (\simm$\pm$1\,kpc), effectively dominating the uncertainty of the distance estimates.
In Fig.\,\ref{fig:rgal_hist} we show the histogram of Galactocentric (upper panel) and heliocentric (lower panel) distances for all velocity components above the $5\sigma$\,rms noise level. The peaks are not only tracing the spiral arms, but also result from complexes lying at similar Galactocentric distances. This can also be seen in Fig.\,\ref{fig:LV}, where the 4 peaks are marked by the coloured dashed lines in the same way as in Fig.\,\ref{fig:rgal_hist}. We find that the peak located at $r_\mathrm{gal}\sim9$\,kpc traces local gas including emission from the Vela molecular ridge. The peak at \simm11\,kpc would trace the Perseus arm, if it was present in the area (see discussion in Sect.\,\ref{sect:structures}). The peaks around $r_\mathrm{gal}\sim12.5$\,kpc (cyan dashed line) trace emission not associated with a spiral arm, being dominated by the huge complex at \elll\simm242\degree\ at $\varv_\mathrm{lsr}$\simm70\,\kms\ (i.e. behind the supershell; see Section\,\ref{sect:outer:co:shell}). The peak found at $r_\mathrm{gal}\sim14$\,kpc partially arises from emission from the Outer arm as well as from structures between the Perseus and Outer arm located at \elll\simm255\degree. Similarly, we see a number of peaks in the heliocentric distance histogram, but due to the projection these cannot be assigned to the arms, as the arms are spread out over a larger heliocentric distance range between $225\degree \leq \ell \leq 260\degree$ than with regard to Galactocentric distance.
\section{Physical Properties}
\label{sec:analysis}\label{sec:properties}
We are now going to investigate the physical properties that can be derived from archival mid-infrared to sub-millimetre continuum emission data in combination with the distances obtained from our pointed CO observations (Section\,\ref{sec:distances}). First we will discuss how we obtained the dust spectral energy distributions for the outer Galaxy. We will then present the derived physical properties and investigate how consistent they are, followed by a detailed look at the results. Here we will mainly investigate the star formation relations as well as the influence of the main structures present in the outer Galaxy on the physical properties of the observed clumps.
\subsection{Dust spectral energy distributions}
\label{sect:seds}
\label{sect:outer:seds}
\setlength{\tabcolsep}{6pt}
\begin{table*}
\begin{center}
\small
\caption[Source parameters obtained from the SEDs for the outer Galaxy]{Source parameters obtained from the SEDs for the first 15 sources: Galactic longitude $\ell$ and latitude $b$, heliocentric distance $R_\mathrm{hel}$, aperture diameter $D_\mathrm{app}$, linear size $D_\mathrm{lin}$, evolutionary class, dust temperature $T_\mathrm{dust}$, and optical depth $\tau_{350}$. Full table available online at CDS.}
\label{tab:seds}
\begin{tabular}{@{}
c
S[table-format=1.1]
S[table-format=1.1]
S[table-format=1.1]
S[table-format=1.1]
S[table-format=1.1]
c
c
r
@{}}
\hline
{Name} & {$\ell_\mathrm{app}$} & {$b_\mathrm{app}$} & {$R_\mathrm{hel}$} & {$D_\mathrm{app}$} & {$D_\mathrm{lin}$} & {Class} & {$T_\mathrm{dust}$} & \multicolumn{1}{c}{$\tau_{350}$}\\
 & {(deg)} & {(deg)} & {(kpc)} & {(\arcsec)} & {(pc)} &  & {(K)} & {}\\
\hline
\hline
G225.020--00.590 & 225.017 & -0.588 & 1.3 & 88.9 & 0.13 & Quiescent & $10.3\pm0.7$ & $(20.6\pm8.4)\times 10^{-5}$\\
G225.030+00.060 & 225.029 & 0.058 & 1.6 & 68.3 & 0.11 & Protostellar & $15.8\pm0.2$ & $(126.6\pm6.4)\times 10^{-7}$\\
G225.080+00.060 & 225.082 & 0.060 & 1.6 & 63.3 & 0.10 & YSO & $14.6\pm0.6$ & $(41.1\pm7.8)\times 10^{-6}$\\
G225.160--00.830 & 225.160 & -0.835 & 1.7 & 69.6 & 0.12 & YSO & $19.9\pm1.4$ & $(19.7\pm4.9)\times 10^{-6}$\\
G225.160--00.840 & 225.163 & -0.841 & 1.7 & 65.9 & 0.11 & Protostellar & $16.5\pm2.0$ & $(16.0\pm7.9)\times 10^{-6}$\\
G225.170--00.750 & 225.167 & -0.746 & 1.3 & 53.1 & 0.06 & YSO & $16.5\pm1.9$ & $(2.5\pm1.2)\times 10^{-5}$\\
G225.210--01.110 & 225.215 & -1.110 & 1.3 & 200.0 & 0.31 & Quiescent & $12.0\pm0.1$ & $(32.1\pm1.9)\times 10^{-5}$\\
G225.220--01.200 & 225.220 & -1.195 & 1.2 & 59.3 & 0.07 & Quiescent & $17.7\pm0.6$ & $(26.9\pm3.6)\times 10^{-6}$\\
G225.230--00.960 & 225.228 & -0.961 & 1.3 & 89.5 & 0.13 & Quiescent & $16.1\pm0.5$ & $(51.4\pm7.0)\times 10^{-6}$\\
G225.240--01.110 & 225.243 & -1.106 & 1.2 & 42.5 & 0.03 & Quiescent & $15.3\pm0.2$ & $(48.7\pm2.4)\times 10^{-6}$\\
G225.300--01.090 & 225.300 & -1.093 & 1.3 & 55.3 & 0.07 & YSO & $12.7\pm1.8$ & $(10.7\pm7.7)\times 10^{-5}$\\
G225.320--00.280 & 225.319 & -0.277 & 1.3 & 90.5 & 0.13 & Quiescent & $13.1\pm0.6$ & $(25.6\pm5.3)\times 10^{-5}$\\
G225.320--01.100 & 225.323 & -1.103 & 1.3 & 85.7 & 0.12 & Quiescent & $9.7\pm0.9$ & $(4.8\pm3.0)\times 10^{-4}$\\
G225.320--01.170 & 225.315 & -1.170 & 1.2 & 51.8 & 0.05 & Protostellar & $14.8\pm2.0$ & $(11.9\pm7.0)\times 10^{-5}$\\
G225.330--00.540 & 225.330 & -0.535 & 1.6 & 42.5 & 0.04 & YSO & $21.0\pm3.4$ & $(4.9\pm2.6)\times 10^{-4}$\\

\hline
\end{tabular}\end{center}
\end{table*}
\setlength{\tabcolsep}{6pt}
To obtain and fit the dust spectral energy distributions (SEDs) we follow the procedures we described in detail in \citet{Koenig2017} and \citet{Urquhart2018}, hence we only give a brief overview here.
We use archival mid-infrared to sub-millimetre continuum maps to obtain the SEDs in 9 different bands at 8, 12, 14, 21/22, 70, 160, 250, 350 and 500\,\micron. In contrast to our previous work, there is no ATLASGAL data available at 870\,$\mu$m for the outer Galaxy, so we use the 250\,\mum positions and source sizes obtained from the source extraction process as described in Section\,\ref{sect:data} for the photometry. In the far-infrared to submm regime we use Herschel/Hi-GAL \citep{Molinari2010} PACS \citep{Poglitsch2010} and SPIRE \citep{Griffin2010} emission maps, covering the cold dust emission from 70\,$\mu$m to 500\,$\mu$m. This data is complemented by mid-infrared maps from WISE \citep{Wright2010} and MSX \citep{price2001} covering the wavelength regime between 8\,$\mu$m and 22\,$\mu$m that is mostly dominated by the emission of a hot embedded component of (proto-)stars. The fluxes for each band are obtained through aperture photometry as described in detail in our previous work \citep{Urquhart2018}, reconstructing the SED of each source. Whenever a flux was measured in the WISE emission maps at the 12 or 22\,\micron bands, it was preferred over the corresponding fluxes at 12 or 21\,\micron determined for the MSX bands due to the lower noise level and moderately better resolution of the WISE images. In this way we obtained SEDs for all observed positions as well as for off-positions that were matched with the extracted sources.
\begin{figure}[tp!]
   \centering
   \includegraphics[trim=0.6cm 0cm 0cm 0cm, clip, scale=0.9]{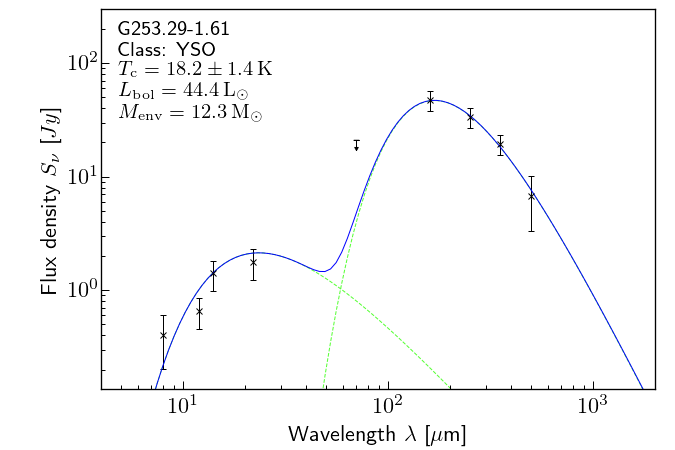}
      \caption{Sample SED (G253.29-1.61) with a hot and cold component, classifying this source as a YSO.}
         \label{fig:sed}
\end{figure}
We have fitted the SEDs using a single component grey-body or two-component model, depending on the emission in the mid-infrared bands. A sample two-component SED is shown in Fig.\,\ref{fig:sed}.  In contrast to our previous work we use the emission found in the 350\,$\mu$m SPIRE band as the reference wavelength due to the absence of a flux measurement at 870\,$\mu$m. In total we were able to fit \varNumSEDsFitted\ SEDs (\simm77\%) of the 791 sources. The SEDs for \varNumSEDsNotFitted\ sources were either not recovered completely due to sensitivity, the source being located in a crowded region, the SED being irregular or the source was too close to the edge of the Hi-GAL area (or a combination of those). We summarize the parameters used to obtain the SEDs and the fitted parameters (dust temperature and opacity) in Table\,\ref{tab:seds} alongside the evolutionary classes as determined from the SEDs (see Section\,\ref{sect:evolution}).
\subsection{Deriving physical properties}
\label{sect:outer:physical_properties}
\setlength{\tabcolsep}{6pt}
\begin{table*}
\begin{center}
\small
\caption[Physical parameters derived from the dust SEDs for the outer Galaxy]{Physical parameters derived from the dust SEDs for the first 15 sources: bolometric luminosity $L$, clump mass $M$, peak column density $N_\mathrm{H_2}$, and luminosity-to-mass ratio $L/M$. Full table available online at CDS.}
\label{tab:seds_physical}
\begin{tabular}{@{}crrrr@{}}
\hline
{Name} & \multicolumn{1}{c}{$L$} & \multicolumn{1}{c}{$M$} & \multicolumn{1}{c}{$N_\mathrm{H_2}$} & \multicolumn{1}{c}{$L/M$} \\
{} & \multicolumn{1}{c}{(\lsun)} & \multicolumn{1}{c}{(\msun)} & \multicolumn{1}{c}{(cm$^{-2}$)} & \multicolumn{1}{c}{(\lsun/\msun)} \\
\hline
\hline
G225.020--00.590 & $(49.5\pm3.2)\times 10^{-2}$ & $(5.3\pm1.7)\times 10^{+0}$ & $(31.8\pm1.6)\times 10^{20}$ & $(9.3\pm3.0)\times 10^{-2}$ \\
G225.030+00.060 & $(383.5\pm4.7)\times 10^{-3}$ & $(31.6\pm8.7)\times 10^{-2}$ & $(22.1\pm1.8)\times 10^{19}$ & $(12.1\pm3.3)\times 10^{-1}$ \\
G225.080+00.060 & $(58.0\pm2.5)\times 10^{-2}$ & $(9.0\pm2.5)\times 10^{-1}$ & $(54.8\pm5.3)\times 10^{19}$ & $(6.4\pm1.8)\times 10^{-1}$ \\
G225.160--00.830 & $(29.1\pm2.1)\times 10^{-1}$ & $(6.2\pm1.7)\times 10^{-1}$ & $(27.6\pm2.4)\times 10^{19}$ & $(4.7\pm1.3)\times 10^{+0}$ \\
G225.160--00.840 & $(56.5\pm6.9)\times 10^{-2}$ & $(3.6\pm1.1)\times 10^{-1}$ & $(28.0\pm2.9)\times 10^{19}$ & $(15.8\pm5.1)\times 10^{-1}$ \\
G225.170--00.750 & $(64.5\pm7.6)\times 10^{-2}$ & $(24.4\pm8.0)\times 10^{-2}$ & $(33.8\pm5.7)\times 10^{19}$ & $(26.4\pm9.2)\times 10^{-1}$ \\
G225.210--01.110 & $(97.8\pm1.1)\times 10^{-1}$ & $(4.4\pm1.3)\times 10^{+1}$ & $(146.0\pm1.4)\times 10^{20}$ & $(22.4\pm6.8)\times 10^{-2}$ \\
G225.220--01.200 & $(56.7\pm2.0)\times 10^{-2}$ & $(28.7\pm9.2)\times 10^{-2}$ & $(34.6\pm3.8)\times 10^{19}$ & $(19.8\pm6.4)\times 10^{-1}$ \\
G225.230--00.960 & $(166.7\pm5.5)\times 10^{-2}$ & $(14.9\pm4.6)\times 10^{-1}$ & $(79.0\pm3.8)\times 10^{19}$ & $(11.2\pm3.5)\times 10^{-1}$ \\
G225.240--01.110 & $(223.8\pm2.6)\times 10^{-3}$ & $(26.0\pm8.4)\times 10^{-2}$ & $(6.1\pm1.5)\times 10^{20}$ & $(8.6\pm2.8)\times 10^{-1}$ \\
G225.300--01.090 & $(37.7\pm5.3)\times 10^{-2}$ & $(11.9\pm4.0)\times 10^{-1}$ & $(15.0\pm2.3)\times 10^{20}$ & $(3.2\pm1.2)\times 10^{-1}$ \\
G225.320--00.280 & $(75.1\pm3.2)\times 10^{-1}$ & $(7.3\pm2.3)\times 10^{+0}$ & $(40.6\pm2.0)\times 10^{20}$ & $(10.2\pm3.2)\times 10^{-1}$ \\
G225.320--01.100 & $(17.3\pm1.6)\times 10^{+0}$ & $(11.4\pm3.6)\times 10^{+0}$ & $(71.9\pm4.2)\times 10^{20}$ & $(15.1\pm5.0)\times 10^{-1}$ \\
G225.320--01.170 & $(7.9\pm1.1)\times 10^{-1}$ & $(8.4\pm2.9)\times 10^{-1}$ & $(15.1\pm2.7)\times 10^{20}$ & $(9.4\pm3.5)\times 10^{-1}$ \\
G225.330--00.540 & $(55.4\pm9.0)\times 10^{+0}$ & $(5.0\pm1.6)\times 10^{+0}$ & $(8.0\pm2.2)\times 10^{21}$ & $(11.0\pm3.9)\times 10^{+0}$ \\

\hline
\end{tabular}\end{center}
\end{table*}
\setlength{\tabcolsep}{6pt}
Using the dust temperatures from the fitted SEDs and assigning distances as determined in Section\,\ref{sec:distances}, we calculate the physical properties from the SEDs as described in \citet{Koenig2017}. However, as the reference wavelength has changed to the $S_{350}$ SPIRE band, slight changes were made on how we calculate the clump mass and H$_2$ column density. 
We obtain the clump mass $M_\mathrm{clump}$ from the integrated 350\,$\mu$m flux density $S_{350}$ as
\begin{eqnarray}
M_\mathrm{clump} = d^2 \cdot \frac{ S_{350}}{B_{350}(T_\mathrm{d})} \cdot \frac{\gamma}{\kappa_{350}},
\label{eq.mass}
\end{eqnarray}
\noindent where $d$ is the distance to the source, $B_{350}(T_\mathrm{d})$ the intensity of a blackbody at 350\,$\mu$m at the cold dust envelope temperature $T_\mathrm{c}$. The dust opacity $\kappa_{350}=1.1$\,cm$^2$g$^{-1}$ at 350\,$\mu$m is calculated as the mean of all dust models from \citet{Ossenkopf1994}, using the dust emissivity index of $\beta=1.75$ used for fitting all SEDs.
As the gas-to-dust ratio $\gamma$ increases with Galactocentric distance, due to the decreasing metallicity in the outer Galaxy we apply a correction factor according to the recently determined trend by \citet[Eq.\,2]{Giannetti2017_gas2dust}:
\begin{eqnarray}
\log \gamma = 0.087 \cdot \frac{R_\mathrm{gal}}{\mathrm{[kpc]}} + 1.44
\end{eqnarray}
Here a linear slope is applied to the logarithm of the gas-to-dust ratio $\gamma$, increasing the factor from $\gamma \sim 150$ near the solar circle to $\gamma \sim 550$ at 15\,kpc Galactocentric radius. We point out that without this correction the clump masses in the outer Galaxy would be underestimated by up to a factor of five compared to masses calculated using the widely adopted value of 100 for the gas-to-dust ratio for the inner Galaxy.
Similarly, we adopt the calculation of the column density $N_\mathrm{H_2}$ to the reference wavelength of 350\,$\mu$m:
\begin{eqnarray}
N_\mathrm{H_2} = \frac{F_{350}}{B_{350}(T_\mathrm{c})} \cdot \frac{\gamma}{\kappa_{350}} \cdot \frac{1}{\Omega_\mathrm{app} \cdot \mu_\mathrm{H_2}\cdot m_\mathrm{H}},
\label{eq.columndensity}
\end{eqnarray}
\noindent were $F_{350}$ is the peak flux density, $\Omega_\mathrm{app}$ being the beam solid angle, and $\mu_\mathrm{H_2}=2.8$ the mean molecular weight of the interstellar medium with respect to a hydrogen molecule \citep{Kauffmann2008} and $m_\mathrm{H}$ the mass of a hydrogen atom. 
 We summarize the physical properties derived from the dust emission for each source in Table\,\ref{tab:seds_physical}. A summary of the dust properties, also taking into account their evolutionary stage (Section\,\ref{sect:evolution}), can be found in Table\,\ref{tbl:derived_para_all}. We will discuss the physical properties in detail and put them into context of their Galactic environment in the following sections.
\subsection{Consistency checks}
\begin{figure}[tp!]
	\centering
	\includegraphics[]{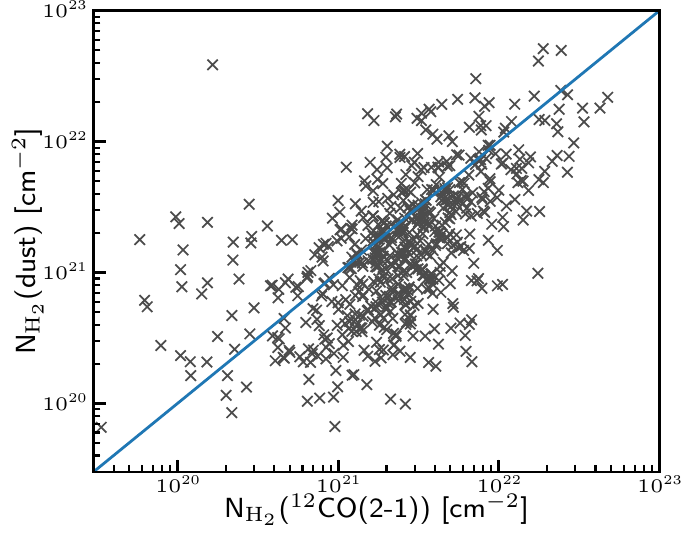}\\
    \includegraphics[]{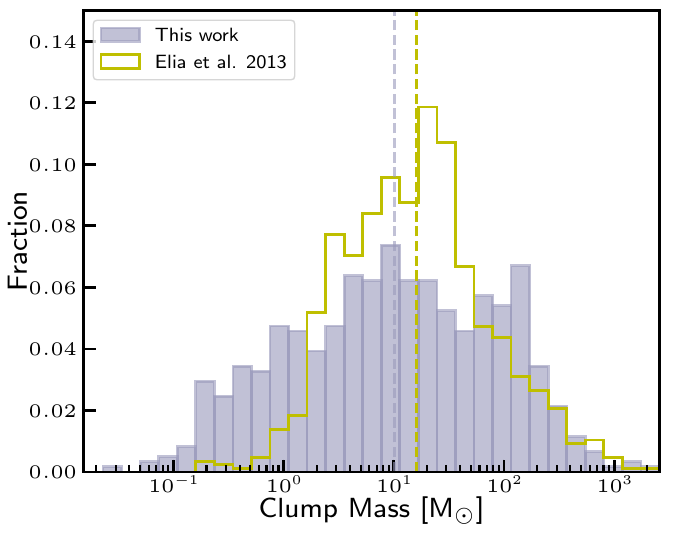}
	\caption[Consistency checks for the outer Galaxy for column density and clump mass]{Consistency checks. Left: Comparing peak N$_{\mathrm{H}_2}$ column densities derived from dust and $^{12}$CO(2--1). Equality of both quantities is marked by the solid line. Right: Clump masses derived from dust continuum emission as calculated for the present work (blue) and from \citet[][yellow outline]{Elia2013a} for $216.5\degree<\ell<225.5\degree$ from the dust SEDs.}
	\label{fig:nh2_dust_12co}
   \label{fig:hist_m_outer}
\end{figure}

To check the consistency of our method, we compare the N$_{\mathrm{H}_2}$ column densities derived from the SEDs with column densities derived from the $^{12}$CO(2--1) emission. To obtain the column densities from $^{12}$CO(2--1), we calculate them as
\begin{eqnarray}
N_{\mathrm{H}_2} = X_\mathrm{^{12}CO(1-0)} \cdot \frac{1}{0.7} \cdot I(\mathrm{^{12}CO(2-1)}),
\end{eqnarray}
\noindent where we use the H$_2$-to-CO conversion factor of $X_\mathrm{^{12}CO(2-1)}=2.3\times10^{20}$\,cm$^{-2}$\,(\kms)$^{-1}$ obtained by \citet{brand1995} for the outer Galaxy, and the integrated line intensity $I(\mathrm{^{12}CO(2-1)})$ as measured from the observed spectra with a line ratio of $^{12}$CO(2--1)/$^{12}$CO(1--0)$=0.7$ \citep{sandstrom2013}. In the upper panel of Fig.\,\ref{fig:nh2_dust_12co} we show the comparison of the column densities derived from dust and CO. As the CO(2--1) column density is derived from emission measured within a beam of 30\arcsec, it traces the denser regions of the more extended clumps with an average angular source size of 62\arcsec, resulting in slightly higher column densities when compared to those derived from the dust emission. Nevertheless, we find both quantities to be in good agreement (p-value of 0.0155) although a large scatter is observed.
We also compare our results to a similar sample of southern outer Galaxy sources from \citet{Elia2013a} obtained for an adjacent area of the southern sky ($216.5\degree<\ell<225.5\degree$). In the lower panel of Fig.\,\ref{fig:hist_m_outer} we compare the clump masses as derived from the Herschel dust continuum emission of both samples. Similar to the masses calculated for our sample, we apply a correction factor for the varying gas-to-dust ratio found by \citet{Giannetti2017_gas2dust} to the masses calculated by \citet{Elia2013a}. We find the distribution to be statistically different (p-value of $5.3\times10^{-4}$) with the sample of the present work picking up significantly more lower mass sources. This is reflected by the mean values to be almost identical with values of $58.2\pm6.0$ and $56.8\pm5.4$\,\msun for our sample and the \citet{Elia2013a} sample, respectively, but the median values to differ by a factor of 1.6 (10.1 and 16.2\,\Msun, respectively). The difference for these two samples is likely caused by a combination of two effects. First, the areas are non-overlapping, so the differences might reflect intrinsic differences of the two areas covered in the outer Galaxy, especially when taking into account the Galactic supershell GSH242-03+37 covered in the present work (see Section\,\ref{sec.supershell}). Furthermore, the distances for the sample by \citet{Elia2013a} were obtained with the NANTEN 4\,m telescope with a beamsize of 2.6\arcmin\ \citep{kim2004}, only allowing to assign distances to the brighter sources within the beam.
From the comparison with the column densities derived from $^{12}$CO as well as with the clump masses from \citet{Elia2013a}, we conclude that our methods can be considered reliable, as either the distributions are similar as shown by an Anderson-Darling test or agree on average within the margin of error.
\subsection{Distance biases}
\label{sect:outer:biases}
Some physical properties suffer from observational distance biases, such as the bolometric luminosity, clump mass and linear source size. These biases are caused by the fixed sensitivity and resolution of the telescope/instrument used. Due to the limited sensitivity, only the brightest and most massive sources can be observed at the farthest distances. Similarly, the limited angular resolution allows us only to observe sources down to this apparent size, which at farther distances translates to larger linear source sizes. To avoid misleading trends that are introduced by these sensitivity and resolution based selection biases, we determine a completeness limit, above which the survey does not suffer from these selection effect out to the given distance.
First we need to distinguish between distance independent parameters and distance dependent parameters. The dust temperature, peak column density and $L/M$ are distance independent, as they are either intrinsic to the sources or cancel out the distance dependence. As mentioned in the last paragraph, the bolometric luminosity and clump mass are directly scaled by the distance squared, and hence are highly dependent on correct distances and are prone to distance dependent observational biases. The same is true for the linear source size, which is linearly dependent on the distance.
\begin{figure}[tp!]
	\centering
	\includegraphics[]{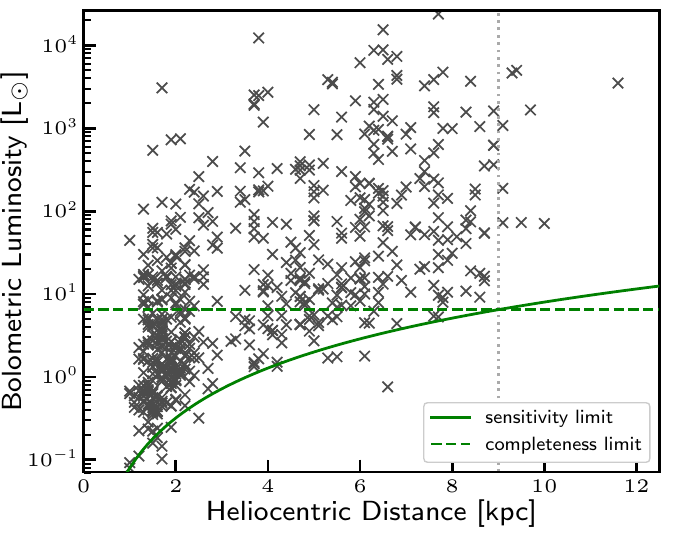}\\
	\includegraphics[]{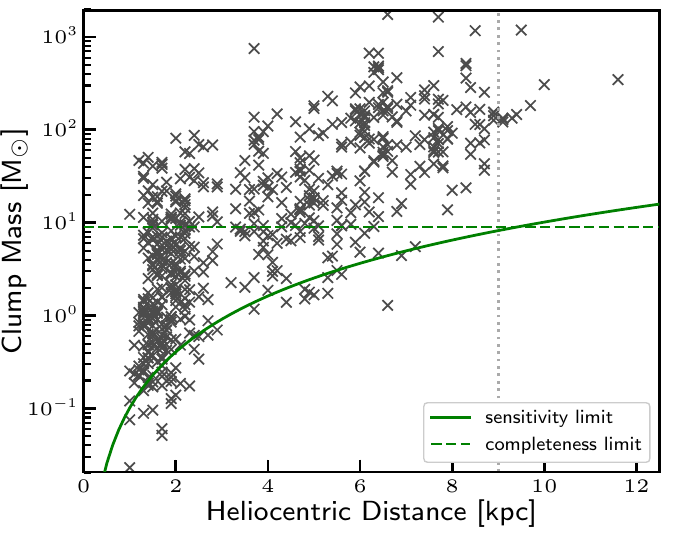}\\
	\includegraphics[]{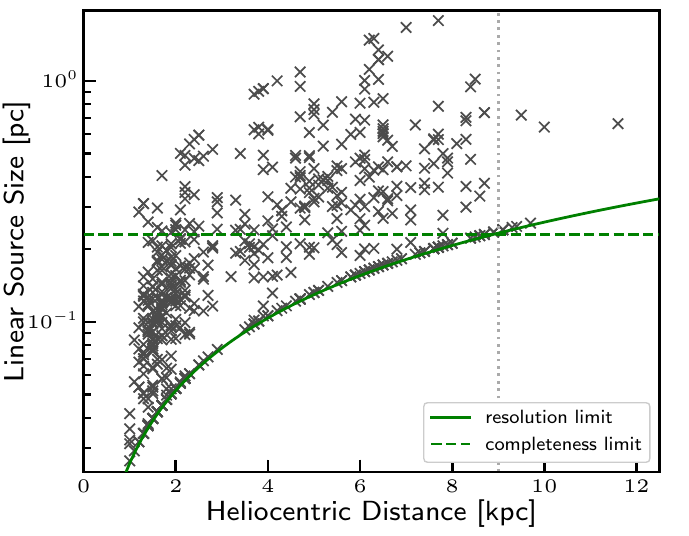}\hspace{7.08cm}
	\caption[Observational distance biases for luminosity, mass and source size]{Observational distance biases. Bolometric luminosity (top-left), clump mass (top right), and linear source size (lower left) versus Heliocentric distance. The solid lines mark the distance dependent sensitivity/resolution limit. The horizontally dashed green lines mark the limit above which our survey is not suffering from a distance bias up to 9\,kpc (vertical dotted line).}
	\label{fig:completeness}
\end{figure}
To determine in which mass and luminosity range our survey is not suffering from distance biases, we show the distribution of luminosities, masses and linear source sizes with respect to heliocentric distance in Fig.\,\ref{fig:completeness}. We calculate the theoretical sensitivity according to Equations\,1 and 2 in \citet{Koenig2017} for the bolometric luminosity and clump mass from the minimum values as input parameters, varying the distance. Similarly we calculate the resolution limit from Equation\,3 in \citet{Koenig2017} but use the beam size for SPIRE\,250 (18.2\arcsec) as input parameter. The theoretical sensitivity and resolution limits are plotted as green solid lines in Fig.\,\ref{fig:completeness}.
As the source count of our sample drops significantly beyond 9\,kpc heliocentric distance we estimate the completeness limit up to this distance as the value of the theoretical sensitivity or resolution limit at 9\,kpc. For the bolometric luminosity, we find that sources out to \simm9\,kpc distance are strongly affected by the sensitivity of this survey for luminosities below 6.5\,\Lsun. Masses of 9\,\Msun\ are found to be the completeness limit for clump mass, whereas for the linear source size, sources smaller than 0.23\,pc suffer from the distance bias. We will filter our samples according to these completeness limits when analysing the distance dependent physical properties.
\section{Discussion}
\label{sect:discussion}
\setlength{\tabcolsep}{6pt}
\begin{table*}[bp!]
\begin{center}
\caption[Summary of physical properties for the outer Galaxy]{Summary of physical properties of the whole population of clumps and the three evolutionary sub-samples identified.}
\label{tbl:derived_para_all}
\small
\begin{tabular}{@{}lc......@{}}
\hline
  \multicolumn{1}{l}{Parameter}&  \multicolumn{1}{c}{\#}&	\multicolumn{1}{c}{$\bar{x}$}  &	\multicolumn{1}{c}{$\frac{\sigma}{\sqrt(N)}$} &\multicolumn{1}{c}{$\sigma$} &	\multicolumn{1}{c}{$x_{\rm{med}}$} & \multicolumn{1}{c}{$x_{\rm{min}}$}& \multicolumn{1}{c}{$x_{\rm{max}}$}\\
\hline \hline
Temperature (K) & 611 & 17.29 & 0.20 & 4.94 & 15.88 & 9.65 & 41.31\\ 
\cline{1-1}
YSO & 259 & 20.18 & 0.35 & 5.62 & 19.10 & 10.57 & 41.31\\ 
Protostellar & 197 & 15.76 & 0.22 & 3.02 & 15.12 & 9.88 & 25.97\\ 
Quiescent & 155 & 14.39 & 0.21 & 2.55 & 14.00 & 9.65 & 24.60\\ 
\hline 

Radius (pc) & 611 & 0.40 & 0.02 & 0.46 & 0.30 & 0.06 & 6.21\\ 
\cline{1-1}
YSO & 259 & 0.49 & 0.03 & 0.55 & 0.39 & 0.07 & 6.21\\ 
Protostellar & 197 & 0.34 & 0.03 & 0.38 & 0.25 & 0.06 & 4.65\\ 
Quiescent & 155 & 0.34 & 0.03 & 0.36 & 0.24 & 0.06 & 3.14\\ 
\hline 

Log[Luminosity (L$_\odot$)] & 611 & 1.16 & 0.04 & 1.10 & 1.03 & -1.10 & 4.38\\ 
\cline{1-1}
YSO & 259 & 1.88 & 0.07 & 1.09 & 1.88 & -0.54 & 4.38\\ 
Protostellar & 197 & 0.75 & 0.05 & 0.75 & 0.71 & -1.03 & 2.39\\ 
Quiescent & 155 & 0.48 & 0.06 & 0.71 & 0.41 & -1.10 & 2.80\\ 
\hline 

Log[Clump Mass (M$_\odot$)] & 611 & 0.99 & 0.04 & 0.93 & 1.00 & -1.64 & 3.24\\ 
\cline{1-1}
YSO & 259 & 1.31 & 0.06 & 0.92 & 1.45 & -1.05 & 3.24\\ 
Protostellar & 197 & 0.78 & 0.06 & 0.89 & 0.79 & -1.22 & 2.71\\ 
Quiescent & 155 & 0.71 & 0.07 & 0.82 & 0.70 & -1.64 & 3.21\\ 
\hline 

Log[$L/M$ (L$_\odot/$M$_\odot$)] & 611 & 0.17 & 0.03 & 0.67 & 0.10 & -1.21 & 2.17\\ 
\cline{1-1}
YSO & 259 & 0.57 & 0.04 & 0.66 & 0.59 & -1.21 & 2.17\\ 
Protostellar & 197 & -0.04 & 0.04 & 0.50 & -0.06 & -1.18 & 1.67\\ 
Quiescent & 155 & -0.23 & 0.04 & 0.47 & -0.26 & -1.19 & 1.44\\ 
\hline 

Log[Peak N$_{\mathrm{H}_2}$ (cm$^{-2}$)] & 611 & 21.26 & 0.02 & 0.55 & 21.28 & 19.80 & 22.61\\ 
\cline{1-1}
YSO & 259 & 21.46 & 0.03 & 0.56 & 21.56 & 19.90 & 22.61\\ 
Protostellar & 197 & 21.12 & 0.04 & 0.50 & 21.12 & 19.92 & 22.39\\ 
Quiescent & 155 & 21.10 & 0.04 & 0.49 & 21.16 & 19.80 & 22.16\\ 
\hline 

\end{tabular}
\end{center}
\end{table*}
\setlength{\tabcolsep}{6pt}
In this section we will discuss the physical properties derived in the previous section. We put a focus on the analysis of the properties found in the outer Galaxy with respect to the structures found here.
\subsection{Evolutionary Sequence}
\label{sect:evolution}
Based on the flux densities obtained for the SEDs we determine the evolutionary state of the sources, following the classification scheme introduced in \citet{Koenig2017}, which was also used by \citet{Urquhart2018} for the entire ATLASGAL sample.
We follow here the naming scheme of our latest paper and refer to mid-infrared bright sources as young-stellar-objects (YSOs). Note however, that these might also include some compact \hii regions, which we did not identify individually. We will therefore refer to the following three classes:
\begin{itemize}
\item Quiescent sources, which are dark at 70\,\mum (no compact emission at 70\,$\mu$m) and represent the earliest phase of star-formation in our sample. These sources might or might not be collapsing.
\item Protostellar sources, which are bright at far-infrared and submm wavelengths but are not sufficiently evolved to produce significant emission at mid-infrared wavelengths ($F_{20}\leq0.1$\,Jy)\footnote{$F_{20}$ is the flux in the 21\,\mum or 22\,\mum band from either MSX or WISE.}. These sources are in the process of collapse and are internally heated.
\item Young-stellar-objects (YSOs) which are bright at mid-infrared wavelengths ($F_{20}>0.1$\,Jy). This group of sources is significantly evolved to produce strong emission at mid-infrared wavelengths by an internal heating source.
\end{itemize}
\begin{figure*}[tp!]
	\centering
	\includegraphics[width=0.38\linewidth]{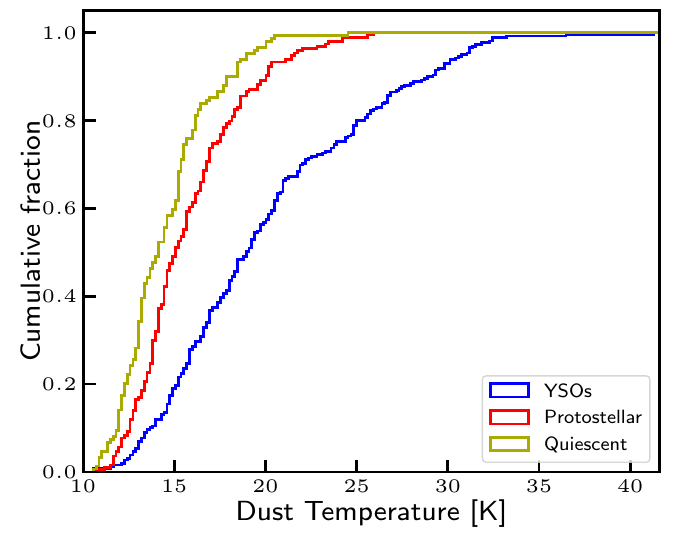}
    \includegraphics[width=0.38\linewidth]{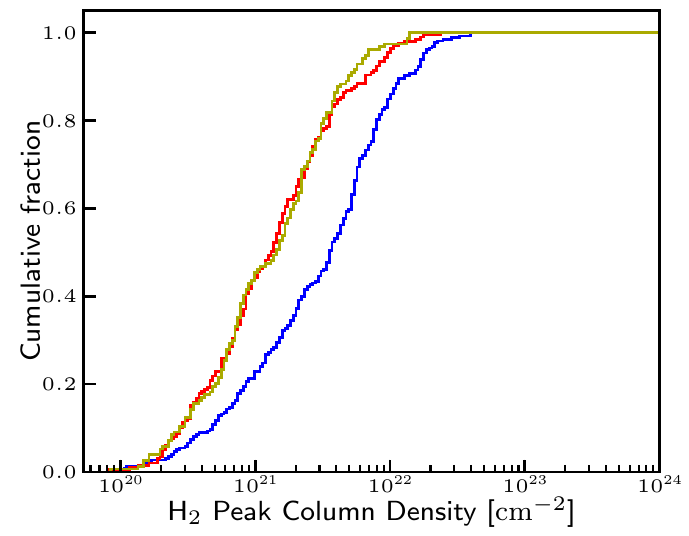}\\
    \includegraphics[width=0.38\linewidth]{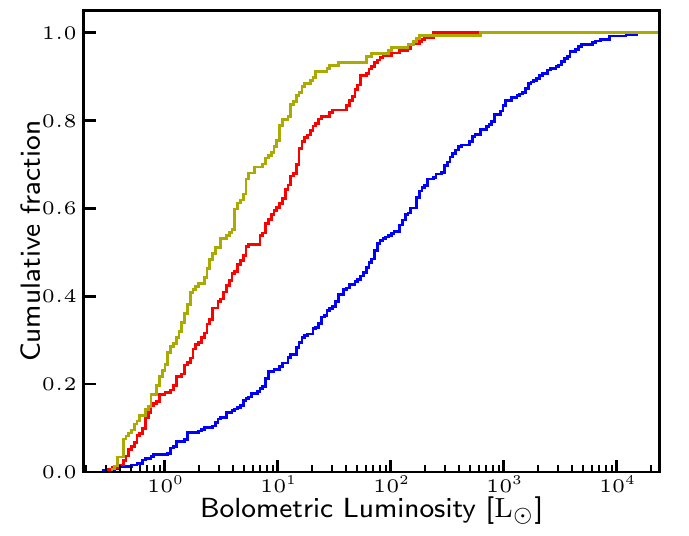}
    \includegraphics[width=0.38\linewidth]{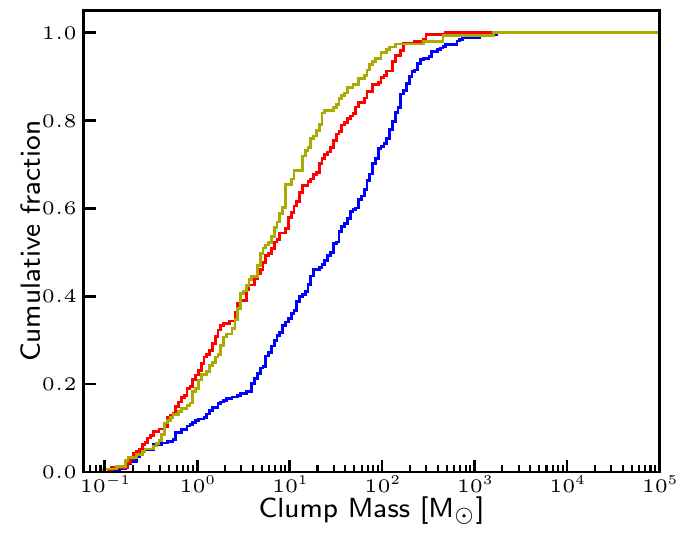}\\
    \includegraphics[width=0.38\linewidth]{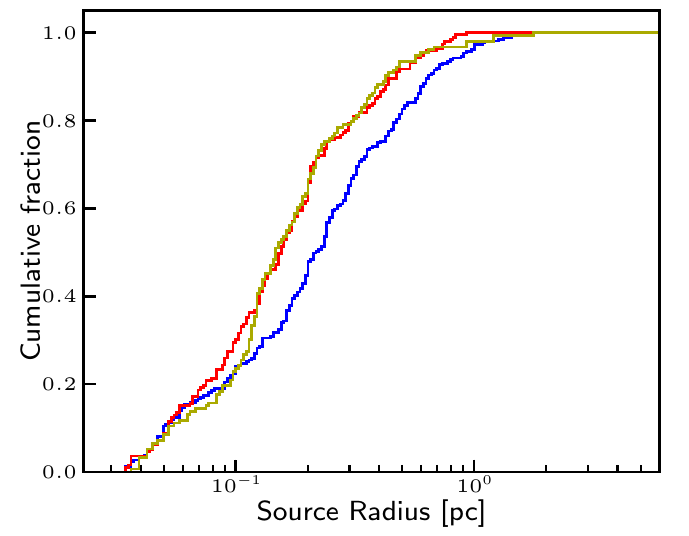}
   \includegraphics[width=0.38\linewidth]{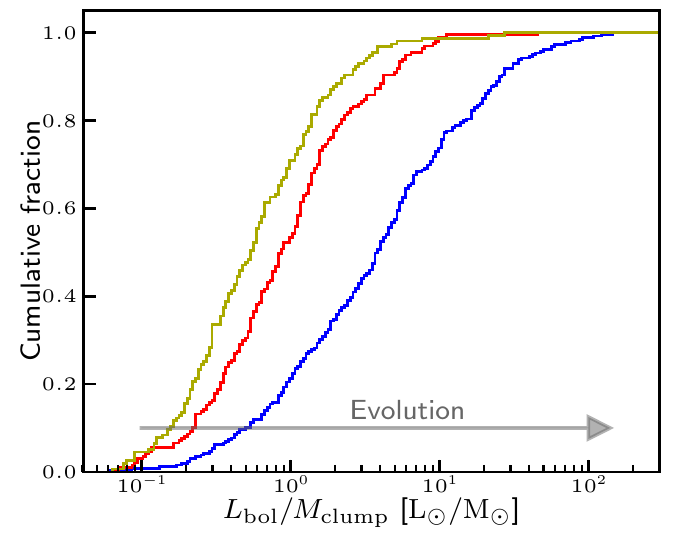}\\
	\caption[Cumulative distributions for derived parameters and evolutionary stages]{Cumulative distributions for the derived parameters of this survey. Coloured lines indicate the distribution for the different evolutionary phases, whereas the dark grey line represents the full sample.}
	\label{fig:sed_hist}
\end{figure*}
This classification scheme has been proven to be reliable as shown by \citet{Giannetti2017_tstruc} using molecular line data. It was further bolstered by our recent work on the ATLASGAL sample \citep{Urquhart2018}, showing clear trends for increasing dust temperature and the bolometric luminosity to clump mass ratio, proving this evolutionary sequence.
Cumulative histograms for the dust properties and the three evolutionary phases of all sources are shown in Fig.\,\ref{fig:sed_hist}. Again, we find the evolutionary stages to be well separated from each other for the dust temperature, the bolometric luminosity and the luminosity-to-mass ratio, as found in \citet{Koenig2017} for the Top100 sample\footnote{The ATLASGAL Top100 sample \citep{Koenig2017} defines a sample of the $approx 100$ brightest sources in 4 distinct evolutionary stages selected from the ATLASGAL survey.}.
This reflects the increase in luminosity due to the accretion of material onto the protocluster and the heating of the local environment as the protostars evolve \citep[compare Sect.\,\ref{sect:star_forming_relations} and][]{Molinari2008}. The similarity between the three evolutionary classes is determined by Anderson-Darling tests, indicating the distributions to be significantly different for each of the aforementioned physical properties, as their pairwise p-values are all well below the threshold of $p<0.0013$, corresponding to a 3$\sigma$ confidence. These three properties (dust temperature, bolometric luminosity and luminosity-to-mass ratio) are therefore excellent indicators of the evolutionary phase of a clump. Care has to be taken though, as there is a large overlap between the different phases, and therefore we can not simply read of the evolutionary phase of a single clump by only taking into account these numbers. However, they can be used to determine statistical properties and identify significant trends in the data.
For the remaining three physical properties (clump mass, linear source size and H$_2$ column density), the quiescent and protostellar evolutionary stages are indistinguishable with a p-value of 0.1. Only the clumps in the YSO phase show, on average, slightly higher values of these properties than the two earlier phases with p-values below $1.8\times10^{-4}$. This indicates that the classification scheme tends to identify the larger, more massive clumps as being more evolved, indicating that the more massive clumps evolve significantly faster than their low-mass counterparts.
\subsection{Physical properties}
\label{sect:outer:temperature}
\paragraph{Temperatures and optical depth:}
Temperature and optical depth are determined as fit parameters from the SEDs and are distance independent. We find that the dust temperatures range from \sedsTMin\,K to \sedsTMax\,K with a mean value of \sedsTMean\,K. We point out that our sample has on average 2\,K lower mean temperature when compared to \citet{Urquhart2018}, probably arising from the difference in sensitivity of the two surveys. We will therefore leave a detailed comparison for \papertwo. For the optical depth we find that all of our sources are optically thin ($\tau \ll 1$) at 350\,$\mu$m, as the optical depths range from \sedsTauMin\ up to a maximum of \sedsTauMax. 
\paragraph{Source sizes:}
\label{sect:outer:linear_source_size}
In Fig.\,\ref{fig:sed_hist}, lower-left panel, we show the distribution of linear source sizes for our sample. Linear source sizes derived from the apparent full-width at half maximum source size vary between \sedsRfwhmMin\,pc and \sedsRfwhmMax\,pc yielding a mean source size of \sedsRfwhmMean\,pc. This is almost identical to the value of \agalRfwhmMean\,pc for the ATLASGAL sample in the 1st and 4th Galactic quadrant \citep{Urquhart2018}, showing that both samples trace structures of similar scale, allowing for a detailed comparison of the different samples.
\paragraph{H$_2$ Column Density:}
The peak H$_2$ column density is found to range between \sedsNHtwopeakMin\,cm$^{-2}$ and \sedsNHtwopeakMax\,cm$^{-2}$ with a mean value of \sedsNHtwopeakMean\,cm$^{-2}$ in the outer Galaxy. As this is below the $10^{23}$\,cm$^{-2}$ found by \citet{Urquhart2018} above which almost all clumps are associated with high mass star formation, we expect to find significantly less massive clumps in the outer Galaxy.
\paragraph{Bolometric luminosity and clump mass:}
The bolometric luminosities for the outer Galaxy sample range between \sedsLbolMin\,L$_\odot$ and \sedsLbolMax\,L$_\odot$ with a mean value of \sedsLbolMean\,L$_\odot$. Clump masses range from \sedsMMin\,M$_\odot$ to \sedsMMax\,M$_\odot$ with a mean value of 58\,M$_\odot$. As with the H$_2$ column density, we expect from these maximum values only a small number of sources to be able to form high mass stars, which we will investigate in the following sections.
\subsection{Star formation relations}
\label{sect:star_forming_relations}
\begin{figure}[tp!]
   \centering
   \includegraphics[scale=1.2]{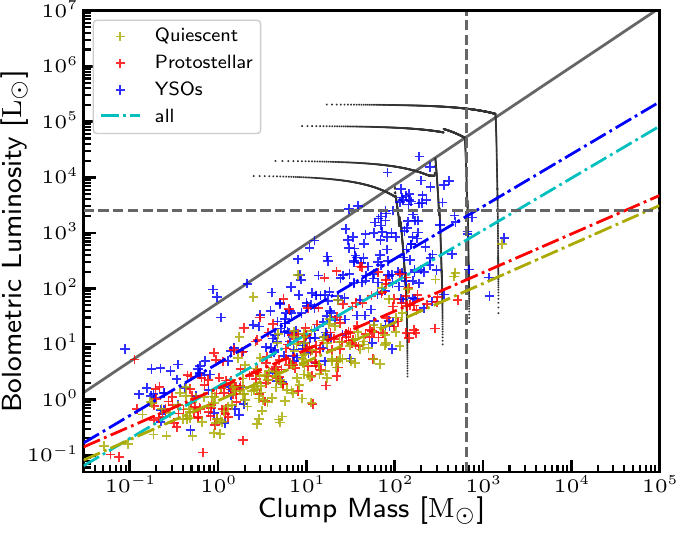}
      \caption[Bolometric luminosity versus clump mass]{Bolometric luminosity versus clump mass. The grey tracks indicate the evolutionary path through the diagram for the given initial masses as calculated by \citet{Molinari2008}. The diagonal solid line marks the expected luminosity for a given mass for a zero age main sequence (ZAMS) star. The horizontal dashed line indicates the expected luminosity for a B2 star ($\sim8$\,\msun) as calculated by \citet{Mottram2011}. The vertical dashed line marks the threshold calculated by \citet{Csengeri2014} above which the clumps are likely to host massive dense cores or a high-mass protostar. The dash-dotted lines are linear fits to the three evolutionary classes.}
   \label{fig:mvsl}
\end{figure}
In Fig.\,\ref{fig:mvsl} we show the bolometric luminosity plotted versus the clump mass. For a given clump one expects the luminosity to increase during the accretion phase, hence moving upwards in the figure until stars reach the zero age main sequence (diagonal solid line). As soon as stars are formed, their feedback disrupts a given clump. As a consequence the clump gets dispersed, decreasing its mass, hence moving left in the diagram. These evolutionary paths through the diagram are indicated by the grey tracks, as calculated by \citet{Molinari2008}. In order to visualize the differences between the evolutionary phases as described in Section\,\ref{sect:evolution}, we show the linear fits to the evolutionary classes as dash-dotted lines. As can be seen, the sources of the different classes move up in the plot as expected from the evolutionary scheme described in Section\,\ref{sect:evolution}. This trend is consistent with the cumulative distribution of the bolometric luminosity-to-clump mass ratio \lsun/\msun\ shown in the lower right panel of Fig.\,\ref{fig:sed_hist}. There we find the three classes to be well separated with all p-values well below 0.0013, following the expected evolutionary trend. 
Further we indicate the expected luminosity for a B2 star ($\sim8$\,\msun) as calculated by \citet{Mottram2011} as dashed horizontal dashed line. Above this threshold we assume that at least one massive star has formed in the clump being responsible for most of its luminosity, but point out that this threshold only sets an upper limit for the number of high mass stars according to our criterion. The vertical dashed line marks the clump mass threshold above which it is likely that the clumps host massive dense cores or a high-mass protostar \citep{Csengeri2014}. From these thresholds we see that only the minority of the sources in the outer Galaxy are able to form a high-mass star (32 sources) and the majority of these have already done so (24 sources at most).
\subsection{Mass-size relation}
\label{sect:mass-size}
\begin{figure}[tp!]
   \centering
   \includegraphics[scale=1.2]{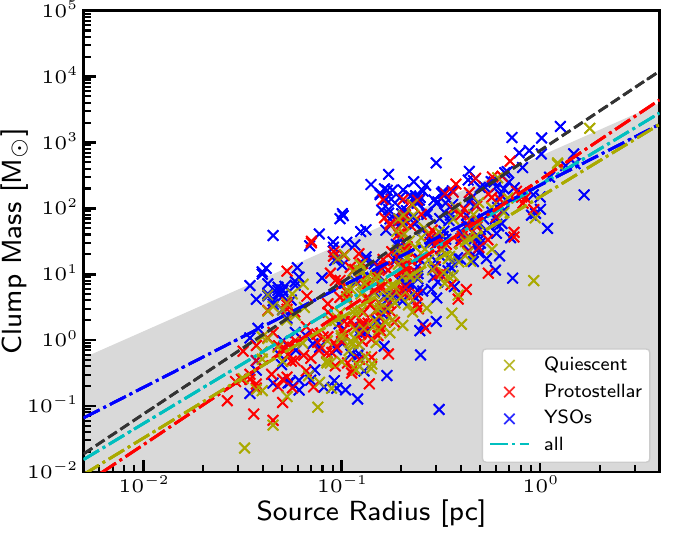}
      \caption[Clump mass versus source radius]{Clump mass versus source radius. The shaded area marks the regime where only low-mass stars form as determined by \citet{Kauffmann2010}. The black dashed line marks the lower limit for effective high-mass star formation as determined by \citet{urquhart2014}, indicating a surface density threshold of 0.05\,g\,cm$^{-2}$. The horizontal grey dotted line marks the threshold of 8\,\msun above which a clump is considered massive. The coloured dash-dotted lines are a linear fit to the corresponding subsamples and the cyan dash-dotted line a fit to the full sample.}
   \label{fig:mvsr}
\end{figure}
In Fig.\,\ref{fig:mvsr} we show the clump mass plotted versus the source radius. The grey dotted line marks the mass-limit for high-mass stars at 8\,\msun. The grey shaded area marks the regime where supposedly only low-mass stars form as determined by \citet{Kauffmann2010}, with only 40 sources (7\%) of the outer Galaxy sample above the threshold and the mass limit. The black dashed line marks the lower limit for effective high-mass star formation as determined by \citet{urquhart2014} indicating a surface density threshold of 0.05\,g\,cm$^{-2}$ found for the inner Galaxy using the ATLASGAL survey. 140 of our sources (i.e. 23\%) are found above this threshold and the mass-limit for high-mass stars. Although we find the majority of the clumps to be able to form only low-mass stars, as we did in the last section, we find a significantly higher fraction to potentially form high-mass stars according to the latter threshold used here. As a consequence we speculate that the threshold for high-mass star formation based on mass surface density is likely to be different in the inner and outer Galaxy, as the radial mass surface density profile of the Galaxy drops significantly according to the review by \citet[][Fig.\,7]{Heyer2015}. 
Finally, we compare the evolutionary phases (colour-coded), but no trend with regard to the evolutionary state of the clumps is found, as can be seen from the linear fits to the individual classes (coloured dash-dotted lines). This is in agreement with our findings in Sect.\,\ref{sect:evolution}.
\subsection{Independent high-mass star formation tracers}
To further identify high-mass star forming clumps, we have also investigated the presence of methanol Class\,\textsc{ii} masers as identified by the Methanol MultiBeam survey \citep[MMB; ][]{green2012_mmb_outer}, which are thought to be exclusively associated with objects in the early phases of high-mass star formation \citep{Urquhart2015}. In the whole area of our survey, only 2 methanol Class\,\textsc{ii} masers are found (G254.880+0.451 and G259.939-0.041), indicating either a lower rate of high-mass star formation, or being the result of the lower metallicity towards the outer Galaxy \citep{Lemasle2018}, and thus reducing the available amount of methanol CH$_3$OH, which contains two metal atoms.
Therefore we also investigated the number of \hii regions in the survey area as identified by the Red MSX Survey \citep{lumsden2013} and the WISE catalogue of Galactic \hii regions \citep{anderson2014}, as the formation of \hii regions is independent of the metallicity. Merging both catalogues we find 10 known \hii regions in our survey area (3 from the RMS survey, 8 from the WISE catalogue with one source in both), as well as another 24 candidate \hii regions from the WISE catalogue. These are about a factor 20--30 less sources per unit-area as compared to the inner Milky Way for the aforementioned surveys.
As these numbers are in the same order as the results from our dust continuum emission in the previous paragraphs, we conclude that there is only very limited high-mass star formation occurring in the outer Galaxy. 
\section{Structure of the outer Galaxy}
\label{sect:structures}
\label{sec:structures}
In this section we use the distance information as derived from the radial velocities to give an overview of the structure of the area investigated in this work. We will briefly investigate the spiral structure as found toward the 3rd Galactic quadrant, and look into the influence of a Galactic supershell known to exist in the observed region.
\subsection{Spiral arms and inter-arm regions}
\label{sect:spiral_arms}
\begin{figure}[!tp]
   \centering
   \includegraphics{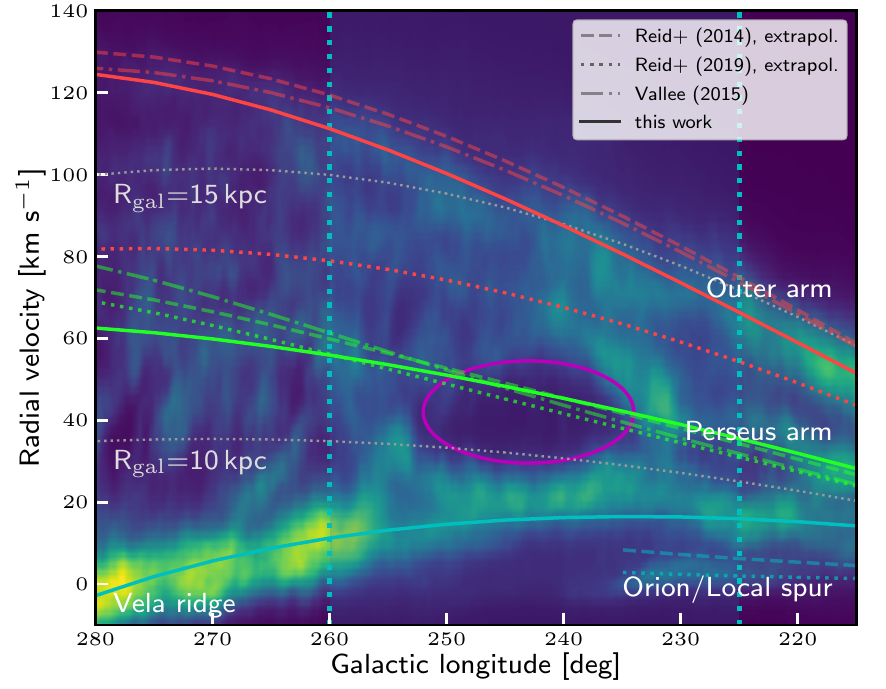}
      \caption{\Hi emission from the Galactic All Sky HI Survey \citep[GASS;][]{McClure-Griffiths2009} for a larger region of the outer Galaxy covering $215\degree<\ell<280\degree$ indicating the integrated brightness temperature between $-5\degree < b < 1.5\degree$. Overplotted are the positions of the spiral arms from \citet{Reid2014}, \citet{Reid2019} and \citet{Vallee2015} as dashed, dotted, and dash-dotted lines, respectively. The spiral arms and local emission as determined in the present work are marked by the solid lines. The dotted grey lines mark Galactocentric radii at 10 and 15\,kpc. the solid magenta ellipse marks the position of the Galactic supershell GSH\,242-3+37. The light blue area framed by the blue dotted lines marks the area of the present work.}
         \label{fig:spiral_arms}
\end{figure}
\begin{figure}[tp!]
   \centering
   \includegraphics[trim=0cm 0.4cm 0cm 0cm, clip, scale=1.0]{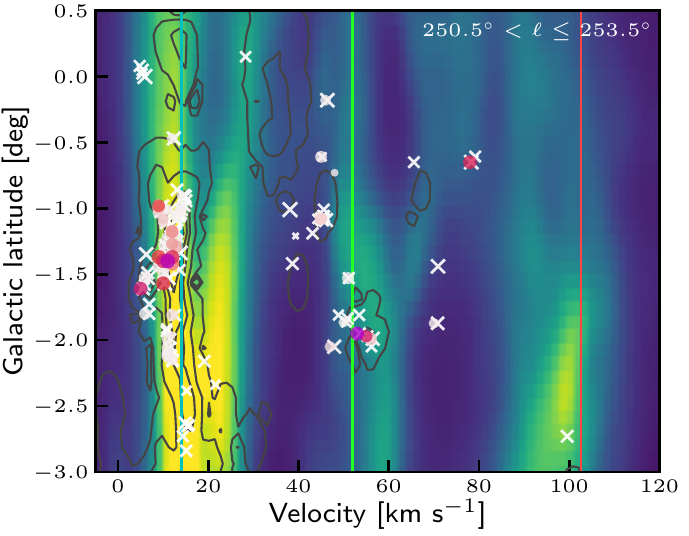}\\
   \includegraphics[trim=0cm 0.4cm 0cm 0cm, clip,scale=1.0]{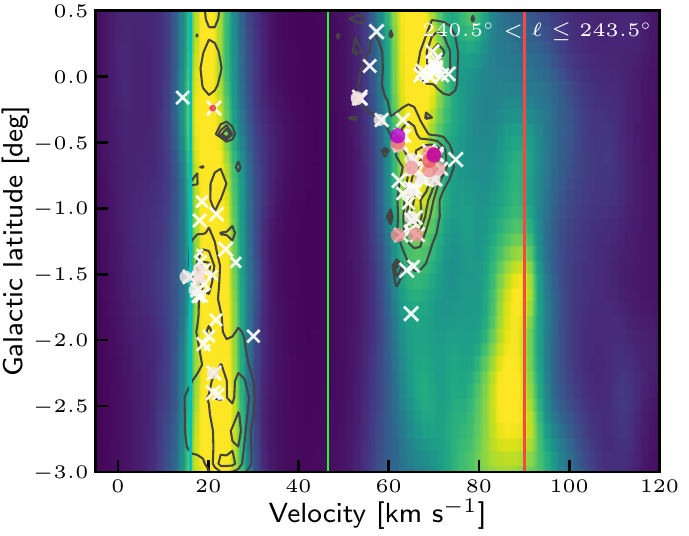}\\
   \includegraphics[scale=1.0]{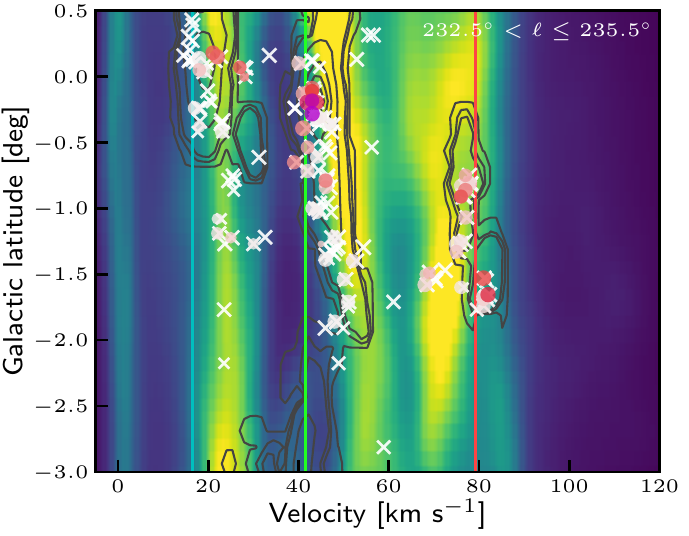}
      \caption[Slices of Galactic latitude vs. radial velocity for the supershell edges and center]{Slices of Galactic latitude vs. radial velocity integrated over $3\degree$ in Galactic longitude at the edges (top, bottom) and the center (middle) of the Galactic supershell. Coloured dots mark the positions of the dust clumps, white crosses mark the positions of clouds identified by the CO(2--1) pointed observations that are not associated with the dust clump. The background image shows the corresponding \hi integrated intensity from the GASS survey \citep{McClure-Griffiths2004}. The contours mark the CO(1--0) emission from \citet{Dame2001} at levels of $5\sigma$, $7\sigma$ and the 10th, 30th, 50th and 70th percentile. The spiral arms as determined in the present work and the position of the local emission are marked by the coloured solid lines. Cyan: local emission; green: Perseus arm; red: Outer arm.}
         \label{fig:HI_bv_shell}
\end{figure}

The nature of the structure of the Milky Way is still under debate \citep{dobbs2014}, and it is unclear weather it is a '\textit{flocculent}' or '\textit{grand-design}' spiral galaxy. The existence of four spiral arms would denominate it as a flocculent galaxy, whereas an interpretation as grand-design would leave the Galaxy with two density wave arms (Perseus and Scutum-Centaurus) and the others being transient features. For this reason, the location and visibility of the spiral arms towards the observed Galactic quadrant is also still under debate with either a single arm or two arms visible \citep[compare e.g.][]{Reid2014,reid2016,koo2017}. Here we will focus on a model where both, the Perseus arm and the Outer arm, are found in the southern outer Galaxy, as the Outer arm is clearly present in the \hi emission (compare Fig.\,\ref{fig:spiral_arms}). However, the location of the arms in the 3rd Galactic quadrant is not well constrained, as their loci are extrapolated from the northern hemisphere, where the measurements of spiral arm positions are better constrained.
\setlength{\tabcolsep}{5pt}
\begin{table}[!bp]
\begin{center}
\small
\caption[Spiral arm parameters]{Spiral arm parameters. See text for a detailed description.}
\begin{tabular}{cccc}
\hline
Name & $R_\mathrm{ref}$ & $\beta_\mathrm{ref}$ & $\Psi$\\
 & (kpc) & (deg) & (deg)\\
\hline
\citet{Reid2014}\\ 
\cline{1-1}
Orion/Local & 8.4 & 8.9 & 12.8\\ 
Perseus & 9.9 & 14.2 & 9.9\\ 
Outer & 13.0 & 18.6 & 13.8\\ 
\hline
\citet{Reid2019}\\ 
\cline{1-1}
Orion/Local & 8.3 & 9.0 & 11.4\\ 
Perseus & 8.9 & 40.0 & 10.3\\ 
Outer & 12.2 & 18 & 3.0\\ 
\hline
\citet{Vallee2015}\\ 
\cline{1-1}
Perseus & 7.0 & 90.0 & 13.0\\ 
Outer & 7.0 & 180.0 & 13.0\\ 
\hline 
This work\\ 
\cline{1-1}
Vela/Orion/Local spur & 9.8 & -0.5 & -24.0\\ 
Perseus & 10.9 & -16.6 & 5.7\\ 
Outer & 14.4 & -23.56 & 15.8\\ 
\hline
\end{tabular}
\label{tab:spirals}
\end{center}
\end{table}
We show a larger region towards the outer Galaxy in Fig.\,\ref{fig:spiral_arms}, covering Galactic longitudes $215\degree<\ell<280\degree$ as mapped in \hi by the Galactic All Sky HI Survey \citep[GASS;][]{McClure-Griffiths2009}. To plot the loci of the spiral arms, we use the equation for a log-periodic spiral as described in \citet{Reid2014}:
\begin{equation}
\ln(R_\mathrm{gal}/R_\mathrm{ref}) = -(\beta-\beta_\mathrm{ref})\tan{\Psi}
\end{equation}
where $R_\mathrm{gal}$ is the Galactocentric radius, $R_\mathrm{ref}$ the Galactocentric radius at the reference Galactocentric azimuth $\beta_\mathrm{ref}$ (with $\beta_\mathrm{ref}=0$ towards the sun, increasing with Galactic longitude), and the pitch angle $\Psi$.
In Fig.\,\ref{fig:spiral_arms} we indicate the extrapolation of the spiral arms and Orion/Local spur as determined by \citet{Reid2014} and \citet{Reid2019} from trigonometric parallax measurements for the northern hemisphere as dashed lines. This includes (from low to high \vlsr) the Orion/Local spur, the Perseus arm and the Outer arm. Furthermore, we plot the two arms visible in the 3rd Galactic quadrant from the 4-arm model as determined as the average from several publications by \citet{Vallee2015} as dash-dotted lines. Although the extrapolations for the Perseus spiral arm are all still in good agreement, the extrapolations for the Outer arm deviate quiet drastically from each other \citep[esp.][]{Reid2019} due to the high uncertainties involved, and fail to trace the highest column density \Hi emissions in the present region.
\begin{figure*}[bp!]
   \centering
   \includegraphics[trim=0cm 0.4cm 0cm 0cm, clip, width=0.44\linewidth]{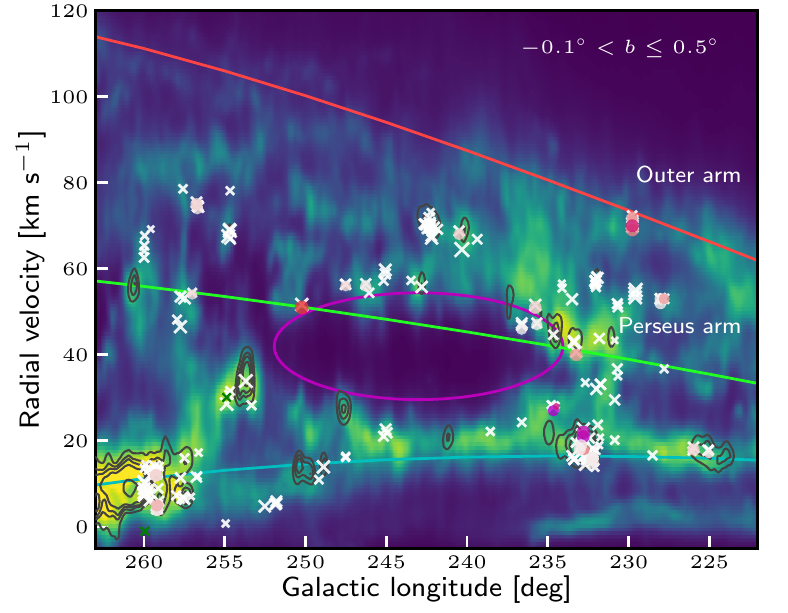}\hspace{0.6cm}\includegraphics[trim=0cm 0.4cm 0cm 0cm, clip, width=0.44\linewidth]{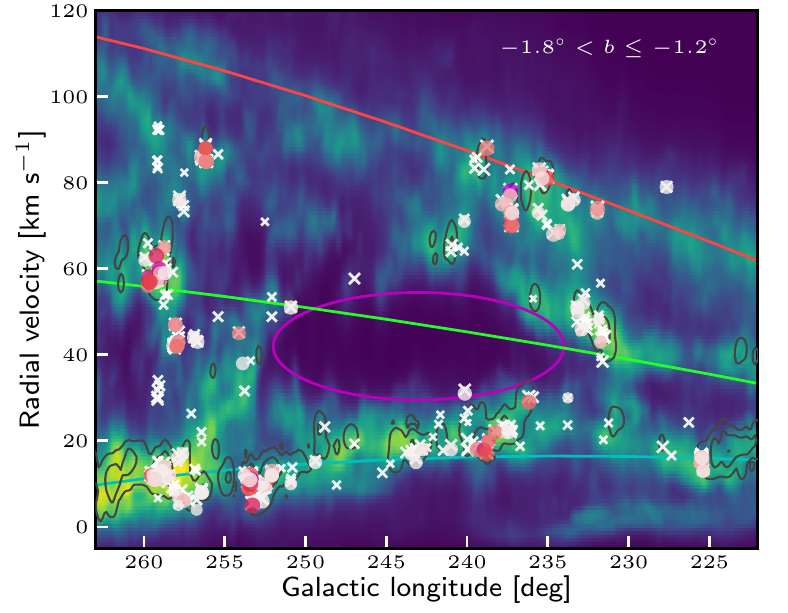}\\
   \includegraphics[trim=0cm 0.4cm 0cm 0cm, clip, width=0.44\linewidth]{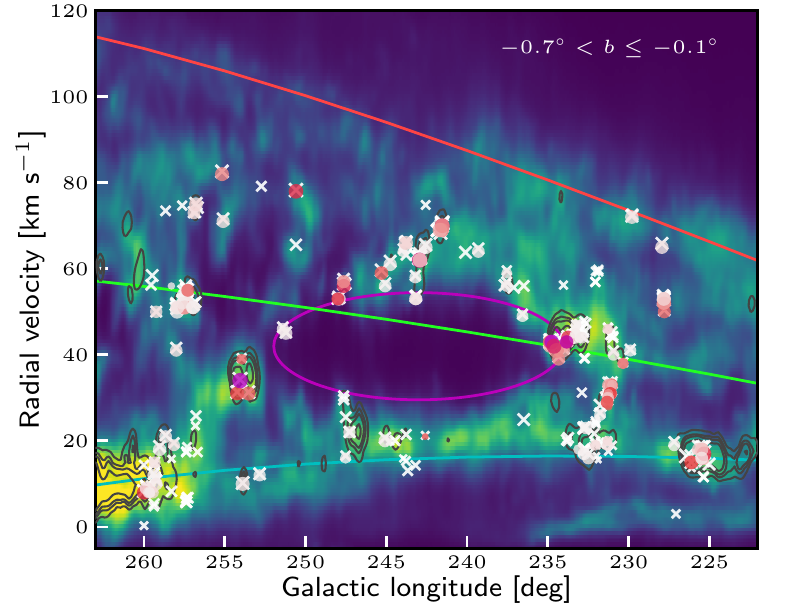}\hspace{0.6cm}\includegraphics[trim=0cm 0.4cm 0cm 0cm, clip, width=0.44\linewidth]{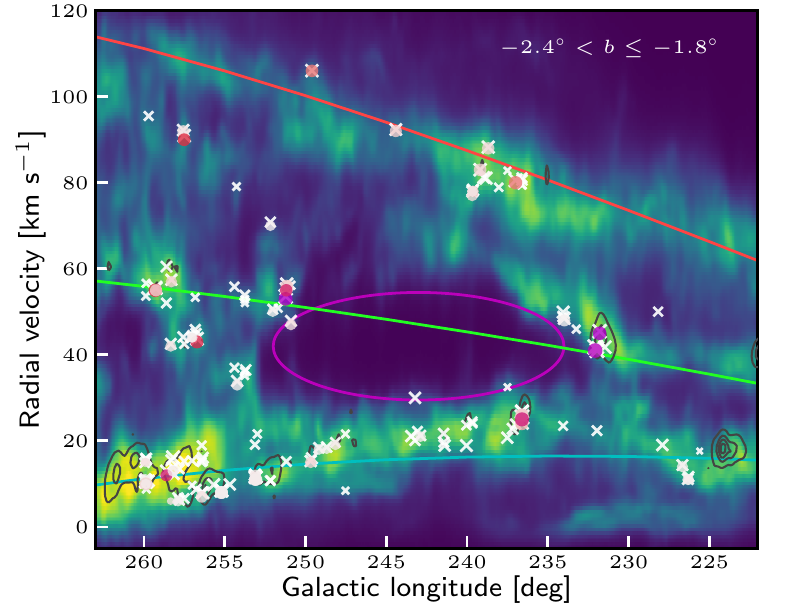}\\
   \includegraphics[width=0.44\linewidth]{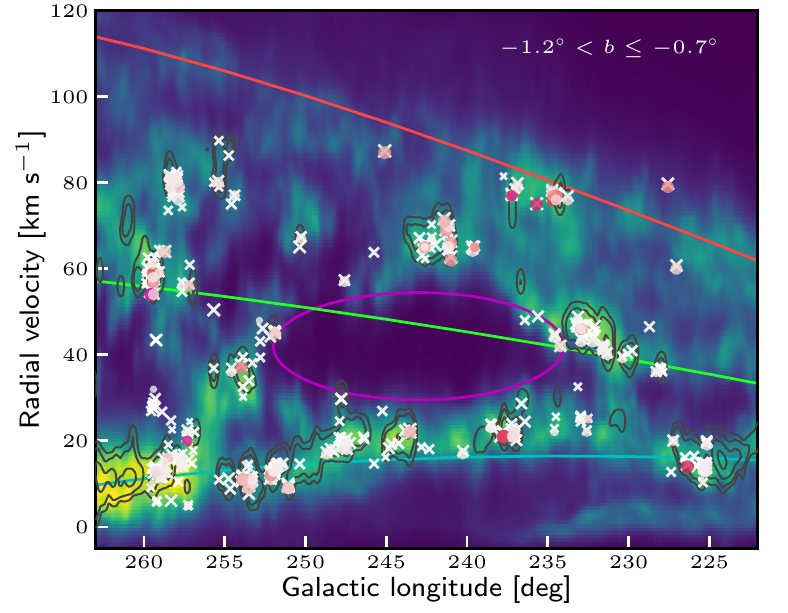}\hspace{0.7cm}\includegraphics[width=0.44\linewidth]{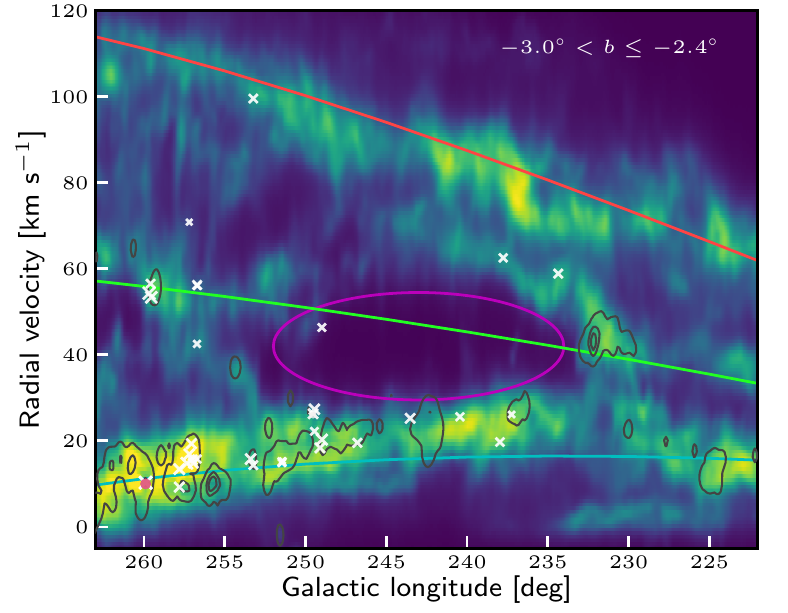}\\
      \caption[Galactic longitude vs. radial velocity for slices of Galactic latitude]{Galactic longitude vs. radial velocity integrated for slices of Galactic latitude. Latitude ranges from $b=+0.5\degree$ (top left panel) to $b=-3.0\degree$ (lower right panel). Crosses mark the CO(2--1) velocity components along a line-of-sight, with coloured dots indicating associated dust clumps. Sizes correspond to the integrated line intensity and the colours for the dust clumps to the 22\,\mum WISE emission being indicative of ongoing star formation. The background image shows the corresponding \hi integrated intensity from the GASS survey \citep{McClure-Griffiths2004}. The contours mark the CO(1--0) emission from \citet{Dame2001} at levels of $3\sigma$, $5\sigma$, $7\sigma$ and the 10th, 30th, 50th and 70th percentile. The spiral arms as determined in the present work and the position of the local emission are marked by the coloured solid lines. Cyan: local emission; green: Perseus arm; red: Outer arm. The solid magenta ellipse marks the position of the Galactic supershell GSH\,242-3+37.}
         \label{fig:HI}
\end{figure*}
Therefore we also plot our own estimates as solid lines for the 3 main features (from low \vlsr\ to high \vlsr): local emission from the Vela molecular ridge (lower left) to the Orion/Local spur (lower right), the Perseus arm (middle) and the Outer arm (upper). As a detailed analysis of the position of the spiral arms is out of the scope of this work, we have manually modified the spiral arm parameters such as to visually match the densest emission as seen in the \Hi integrated intensity map.
Note that the local emission (Vela molecular ridge and Orion/Local spur) is an inter-arm feature between the Sagitarius spiral arm located in the inner Galaxy at $R_\mathrm{gal}\sim7$\,kpc and $\ell=0\degree$ (not visible here) and the outer Galaxy. The position indicated here by the solid cyan line therefore does not represent an arm that is part of the spiral pattern but just a fit to the local emission. In case of the Vela molecular ridge the emission distribution bends inwards with increasing Galactic azimuth with a large negative pitch angle, in contrast to the spiral arms extending outwards with increasing Galactic azimuth and a comparably smaller pitch angle. The parameters used for the spiral arms and the local emission are summarized in Table\,\ref{tab:spirals}.
\begin{figure*}[tp!]
   \centering
   \includegraphics[trim=0cm 0.4cm 0cm 0cm, clip,width=0.39\linewidth]{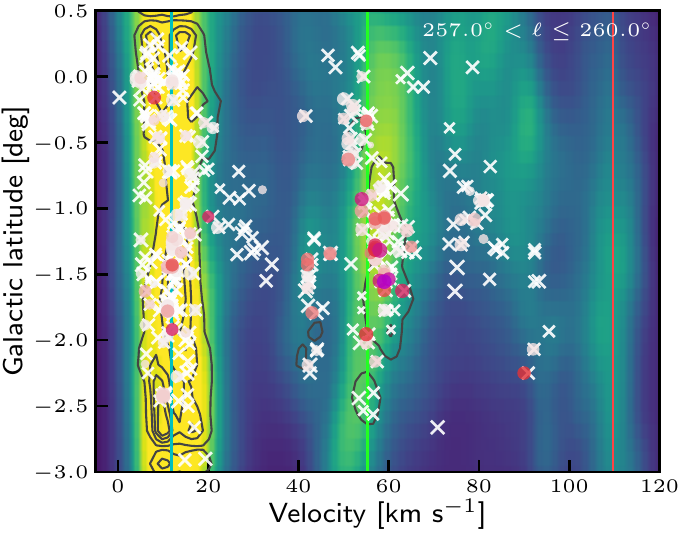}\hspace{0.7cm}\includegraphics[trim=0cm 0.4cm 0cm 0cm, clip,width=0.39\linewidth]{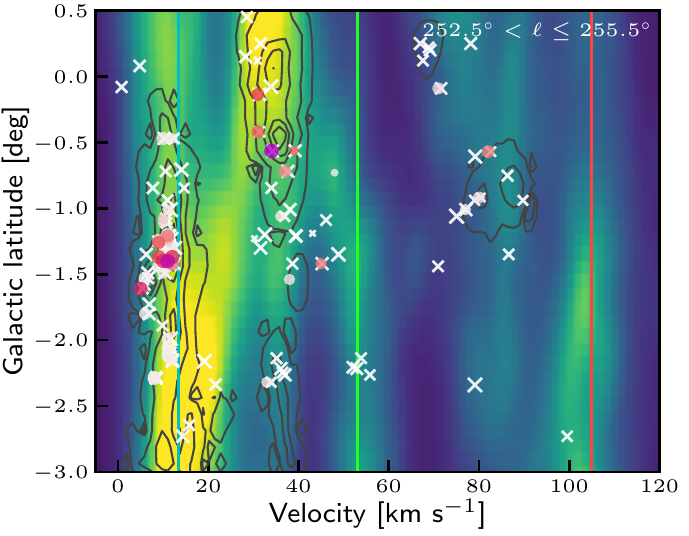}\\
   \includegraphics[width=0.39\linewidth]{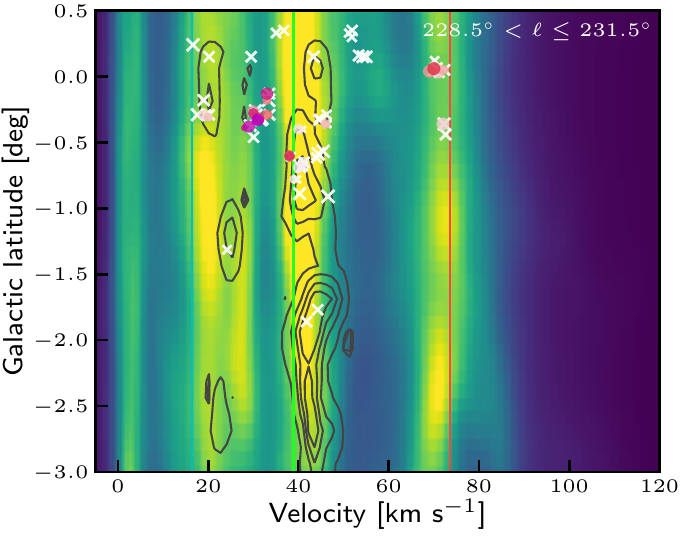}\hspace{0.7cm}\includegraphics[width=0.39\linewidth]{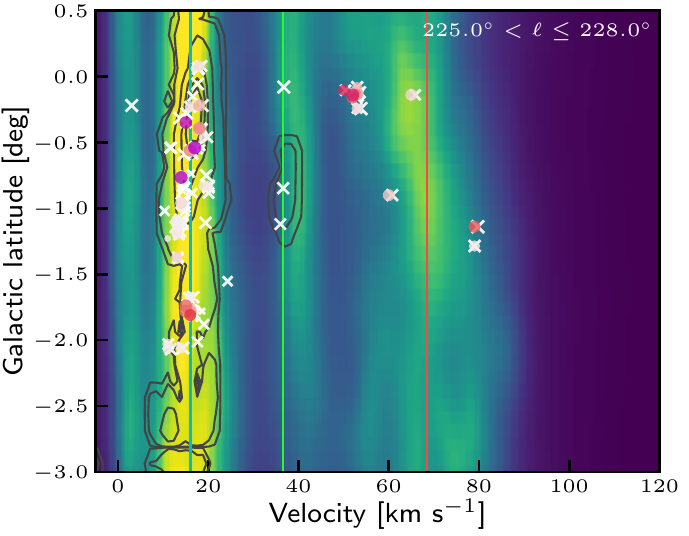}
      \caption[Slices of Galactic latitude vs. radial velocity away from supershell]{Selected slices of Galactic latitude vs. radial velocity integrated over $3\degree$ in Galactic longitude away from the Galactic supershell (top: higher longitudes than shell, bottom: lower longitudes of shell). Crosses mark the CO(2--1) velocity components along a line-of-sight, with coloured dots indicating associated dust clumps. Sizes correspond to the integrated line intensity and the colours for the dust clumps to the 22\,\mum WISE emission being indicative of ongoing star formation. The background image shows the corresponding \hi integrated intensity from the GASS survey \citep{McClure-Griffiths2004}. The contours mark the CO(1--0) emission from \citet{Dame2001} at levels of $5\sigma$, $7\sigma$ and the 10th, 30th, 50th and 70th percentile. The spiral arms as determined in the present work and the position of the local emission are marked by the coloured solid lines. Cyan: local emission; green: Perseus arm; red: Outer arm.}
         \label{fig:HI_bv}
\end{figure*}
\label{sect:interarm}
To give an overview and explore the three dimensional structure of the observed region, we present longitude-velocity ($\ell-v$) plots for different slices of Galactic latitude in Fig.\,\ref{fig:HI} and latitude-velocity ($b-v$) plots for different slices of Galactic longitude in Figs.\,\ref{fig:HI_bv} and \ref{fig:HI_bv_shell}. We show the \hi integrated intensity as a background image \citep{McClure-Griffiths2009}, the CO(1--0) emission from \citet{Dame2001} as black contours, and mark the positions of our sources observed in CO(2--1) as small crosses with those associated with dust clumps as coloured dots. We will now discuss the different features present in the area.
We find that our sources as detected in $^{12}$CO(2--1) are mostly well correlated with the brightest features of the H\textsc{i} emission. They are tracing the local emission very well, but except for the region of the brightest \hi emission located at $\ell\sim233\degree$, are poorly tracing the Perseus arm. We conclude that this is not a sensitivity issue, as plenty of sources between \vlsr\simm80-90\,\kms\ are well detected in the $\ell\sim258\degree$ region, and therefore we find no reason why such sources should not be detected in the Perseus arm between $225\degree<\ell<250\degree$. Similarly, we find the CO(2--1) emission to be poorly correlated with the Perseus arm for some slices of longitude ($\ell\sim254\degree$ and $\ell\sim227\degree$, Fig.\,\ref{fig:HI_bv}, right panels). We rather conclude that the poor coherence of the observed source velocities with the locus of the Perseus arm in $\ell-v$ and $b-v$  plots is a consequence of the presence of the expanding Galactic supershell G\,242-03+37, which we will investigate further in Sect.\,\ref{sect:supershell}.
Several larger features can be identified spanning between the local emission and the suspected positions of the spiral arms (compare Fig.\,\ref{fig:HI}): In the region towards higher longitudes than the supershell ($250\degree<\ell<260\degree$) a web-like network of blobs, bridges or spurs can be seen, with large clusters of clumps correlated with the brightest \hi emission. At least four such complexes can be identified by eye from the $\ell-v$ maps ($\ell\sim 254\degree$, $\varv_\mathrm{lsr}\sim38$\,km\,s$^{-1}$; $\ell\sim 257.5\degree$, $\varv_\mathrm{lsr}\sim50$\,km\,s$^{-1}$; $\ell\sim 258\degree$, $\varv_\mathrm{lsr}\sim80$\,km\,s$^{-1}$; $\ell\sim 259\degree$, $\varv_\mathrm{lsr}\sim60$\,km\,s$^{-1}$), with more isolated clumps along the bridges. These bridges are also well visible in the $b-v$ plot around $\ell\sim285.5\degree$ (Fig.\,\ref{fig:HI_bv}, upper right panel). Note that these bridges might not be visible in CO(1--0) the \citep[black contours][]{Dame2001}, as the sensitivity is rather poor (rms noise up to 0.4\,K). 
Between $225\degree<\ell<240\degree$\ (i.e. lower longitudes than the supershell) we find at the lower latitudes (see right hand column of Fig.\,\ref{fig:HI}) a feature in \hi that is aligned perpendicular to the line-of-sight, indicating the continuation of the Perseus arm from the north. But for higher latitudes (left hand column of Fig.\,\ref{fig:HI}) this structure becomes more complex with a network of bridges spanning between the Orion/Local spur and the Outer arm more parallel to the line-of-sight (and perpendicular to the arms). This can also be seen in the $b-v$ plot in Fig.\,\ref{fig:HI_bv}, lower left panel, where the CO emission at $b\sim0\degree$ is spanning from $\varv_\mathrm{lsr}\sim20$\,\kms\ out to $\varv_\mathrm{lsr}\sim60$\,\kms, whereas the local emission and the Perseus and Outer arm are more clearly separated at lower latitudes. This change in morphology and orientation in dependence on the Galactic latitude could be either interpreted as the Perseus arm being disrupted or this structure not being a spiral arm at all. In contrast to the web of structures located towards higher longitudes than the supershell between 250\degree\ and 260\degree, we only find one larger star forming complex in this area ($\ell\sim 232\degree$, $\varv_\mathrm{lsr}\sim50$\,km\,s$^{-1}$), located at the rim of the supershell (Fig.\,\ref{fig:HI_bv_shell}).
In general, we find the three-dimensional structure of the Galactic disk towards the outer Galaxy in the third quadrant to be rather complex; not only with respect to longitude and velocity/distance, but also with respect to the vertical structure of the thin disc. The structures found in the observed region change rather dramatically with Galactic latitude, showing that the thin disk is a complex three-dimensional web-like structure than a flat pan-cake-like structure.
\subsection{The Galactic supershell GSH\,242--03+37}
\label{sect:supershell}
\label{sect:outer:co:shell}
\label{sec.supershell}
\begin{figure}[tp!]
	\centering
	\includegraphics[]{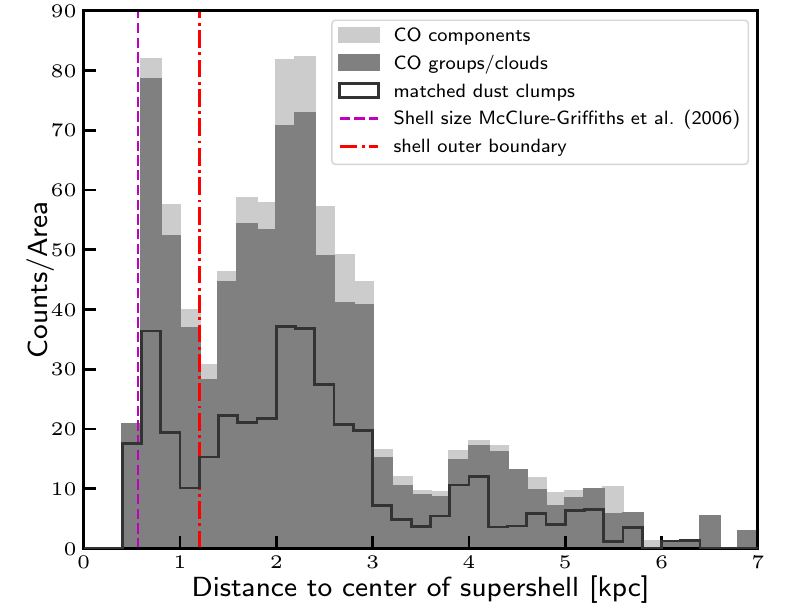}
	\caption{Histogram for the distance of all clouds identified in CO to the centre of the supershell. The edge of the supershell is clearly distinguishable from the other sources, with the central parts at $R<0.4$\,kpc being void of any CO clouds.}
	\label{fig:outer:shell_hist_co}
\end{figure}
One of the most striking features towards the observed region is the void located at $\ell\sim 242\degree$ and \vlsr\,$\sim 40$\,\kms. This structure is known to be the Galactic supershell GSH\,242--03+37 and was first identified by \citet{Heiles1979} and more recently investigated by \citet{McClure-Griffiths2006}. It has a diameter of the order of a kiloparsec and an expansion velocity of $\sim 7$\,\kms. With the shell spanning about 15\degree\ in Galactic longitude, 25\,\kms\ in velocity and centred $-1.6$\degree\ below the Galactic plane, we find that our observed region spans roughly a band around the equator of that supershell. 
As this structure has a profound impact on this region, we will briefly discuss the possible origin of this supershell. \citet{McClure-Griffiths2006} calculate the energy needed to form such an expanding supershell to be in the order of $10^{53}$\,ergs, which is about two orders of magnitude higher than the expected energy input from a single Type\,\textsc{ii} supernova \citep{Woltjer1974}. This would therefore require in the order of hundreds of Type\,\textsc{ii} supernovae to drive this shell, which is rather unlikely, especially as there is no strong X-ray source at the centre of the shell, which would be expected in the presence of hundreds of expected supernova remnants. A more suitable explanation might be the passage of a high velocity cloud (HVC) through the Galactic disk, like the one observed by \citet{Park2016} causing the Galactic shell GS040.2+00.6-70 in the northern hemisphere. Furthermore, HVCs are known to punch holes through galactic discs \citep[e.g.][]{Schulman1996,Boomsma2008}. Preliminary literature research further bolsters this hypothesis \citep[compare e.g.][]{Galyardt2016,McClure-Griffiths2006,PlanckCollaboration2015}, but as an investigation of the origin of this shell is out of the scope of this work, we will instead continue to discuss its impact on the region.
A consequence of the existence of this huge expanding supershell is a reduced reliability of kinematic distances for sources within its vicinity, as the expansion of the shell at a velocity of $v_\mathrm{exp}\sim7$\,\kms\ \citep{McClure-Griffiths2006} distorts the Galactic rotation curve. For the distances to the shells center this mostly affects the sources at the front and rear wall of the supershell, as the line-of-sight velocity \vlsr\ is parallel to the expansion velocity $v_\mathrm{exp}$. However, the expansion velocity $v_\mathrm{exp}$ has no effect on the sources on the northern or southern edges of the shell, as their position relative to the center of the shell is measured from the longitude and the line-of-sight velocity \vlsr\ is not affected as it is perpendicular to the expansion velocity. The effect can be seen in Fig.\,\ref{fig:distr} as the supposedly circular supershell is deformed into an ellipse along the line-of-sight. Accordingly, this needs to be taken into account when interpreting structures and especially spiral arms in this region of the Milky Way. Furthermore, the supershell is coincident with the extrapolation of the Perseus arm, which might be the main reason why the Perseus arm cannot be traced clearly in this part of the outer Galaxy. If a large high-velocity cloud has hit the Perseus arm in the past, such an event might simply have lead to its local destruction, pushing the material of the arm away. Indeed, we see a ring of sources identified in CO emission around this supershell.
\begin{figure*}[!tp]
	\centering
    \includegraphics[width=0.38\linewidth]{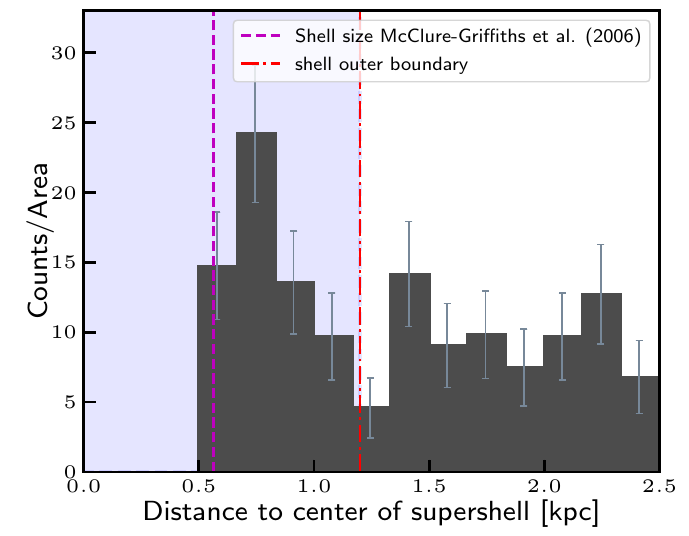}
    \includegraphics[width=0.38\linewidth]{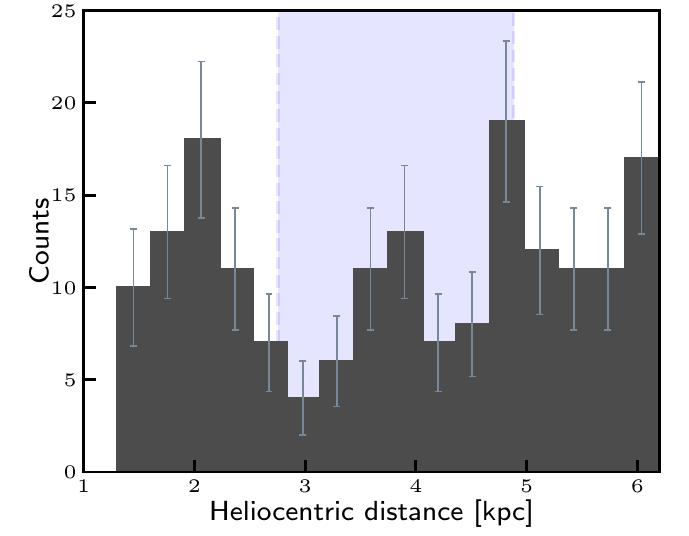}
	\caption[Histograms of distance to the centre of the supershell]{Histograms of the sensitivity filtered sources (see text) for the distance to the centre of the supershell (left) and Heliocentric distance of this subsample. The blue shaded area marks the distance range spanned by the shell.}
	\label{fig:outer:shell_hist_seds}
\end{figure*}
In Fig.\,\ref{fig:outer:shell_hist_co} we present a histogram of the number of CO emission components, clouds and associated dust clumps vs. the centre of the supershell normalized by the area. With the Galactic plane being coincident with the rim around the equator GSH\,242--03+37, we find the central region ($r_\mathrm{GSH}\leq0.4$\,kpc) void of any clumps. However, we find that the number-density of sources is increased in the wall of this supershell between 0.6\,kpc$\leq r_\mathrm{GSH} \leq$1.2\,kpc, which we estimate to be \simm0.6\,kpc wide. In fact, we find a sharp increase in number-density of clumps from the inner void to the wall, which then gradually falls off again\footnote{The broad peak between 1.4 and 3.2\,kpc in Fig.\,\ref{fig:outer:shell_hist_co} is not associated with the shell, but the superposition of different clusters of clumps in the observed region.}. This is in agreement with \citet{McClure-Griffiths2006}, finding a sharp increase in H\textsc{i} from the inner region to the wall of the shell, interpreting it as an indication of compression and being associated with a shock.
\subsection{Physical Properties with respect to the Galactic Supershell}
\label{sect:outer:properties:shell}
In the previous section we had a look at the distribution of clumps around the supershell. In this section we will take a look at the impact of the supershell on its environment.
To avoid observational and selection biases, we limit our sample to only include sources that are symmetrically distributed around the shell with a maximum distance of 2.5\,kpc to its centre, reducing our sample size to 429 sources. Furthermore, we need to filter our sample for sensitivity and resolution thresholds, with the farthest source in this sub-sample is located at 6.2\,kpc Heliocentric distance. For this distance we limit our sample to sources with $L_\mathrm{bol}>2$\,\lsun, $M_\mathrm{clump}>3$\,\msun, and $r_\mathrm{src}>0.15$\,pc according to Fig.\,\ref{fig:completeness}, reducing our sample to 178 sources.
In Fig.\,\ref{fig:outer:shell_hist_seds} we show two histograms for this subsample. First we show the distribution of sources per unit area with respect to the distance to the shells centre (left). Again we are able to identify the central void of the supershell extending up to $\sim0.5$\,kpc as well as the enhanced number of sources located in the shells wall. In the right panel, we show the histogram of heliocentric distances, with the range covered by the supershell marked as blue shaded area. We find no bias to any heliocentric distance, although peaks from local clusters are clearly visible. These are either from local emission (at $\sim2$\,kpc), the edge of the shell (at $\sim4$\,kpc), the rear wall (at $\sim5$\,kpc) or farther out (at 6\,kpc). 
We investigate the influence of the Galactic supershell on the physical properties of the clumps, as well as for the $F_{70\mathrm{ \mu m}}/F_{500\mathrm{ \mu m}}$ flux ratio as an indicator for star-formation activity. For all properties we are unable to find any significant correlation with the distance to the centre of the supershell, with all p-values well below the $3\sigma$ level, accepting the null-hypothesis of the sample being equivalent to a distribution with a slope of zero. Care has to be taken though, as the distances to the shell are determined from kinematic distances, and these are affected by the supershells expansion velocity of \simm7\,\kms\ and thus might have a strong impact on distance dependent properties like the masses, luminosities and linear source sizes. But as this would influence sources in the near and far wall, and thus equally decrease and increase the distances and thus the physical properties, we conclude that the average values in the shells walls are statistically unaffected.
In summary, we find an increase in the number-density of sources in the walls of the supershell, but find their physical properties or the star-formation activity unaffected. In fact, we find the increased number-density of clumps to be consistent with the hypothesis of \citet{McClure-Griffiths2006}, finding that the material in the walls of the supershell might be shocked and compressed. Furthermore, the increased source count per unit area fits perfectly into the picture of \citet{Izumi2014}, who found star formation to be possibly induced by the passage of a high-velocity cloud through the disc in the outer Galaxy in the 2nd Quadrant.
\section{Summary and Outlook}\label{sect:summary}
In order to extend our previous studies of the ISM and star formation \citep{Koenig2017,Urquhart2018} to the outer Galaxy between $225\degree \leq \ell \leq 260\degree$ we used Herschel/Hi-GAL 250\,\mum SPIRE continuum emission maps to select a representative sample of more than 800 sources from a rudimentary source catalogue of more than 25,000 extracted clumps using SExtracor \citep{Bertin1996}, giving positions and source sizes for these clumps. 
We observed these sources in $^{12}$CO(2--1), identifying \varNumGoodCalComplexes\ clouds that consist of a total of \varNumGoodCalComponents\ individual velocity components, for a total of \varNumPos\ positions, including recovered off-positions. \varNumLosSingleComplexes\ (\varNumLosSingleComplexesFraction\%) lines-of-sight were found with a single velocity component, whereas two or more clouds were found towards \varNumLosMultiComplexes\ (\varNumLosMultiComplexesFraction\%) lines-of sight, yielding on average \varNumComplexesPerLoS\ clouds per line of sight. Consecutively, distances were calculated using a rotation model of the Galaxy, applying the rotation curve from \citet{Brand1993} for all clouds and velocity components. For every line-of-sight, we finally associated the cloud with the highest integrated intensity to the according dust clump. Combining our velocity measurements with \Hi emission maps from the GASS survey \citep{McClure-Griffiths2004} and CO(1--0) maps from \citet{Dame2001}, we were able investigate the large-scale structures of the southern outer Galaxy between $225\degree \leq \ell \leq 260\degree$.
We determined physical properties of the selected dust clumps, recovering their dust spectral energy distributions from Hi-GAL, MSX and WISE continuum emission maps. The SEDs were consecutively fitted with a simple two-component model yielding dust temperatures, integrated fluxes and H$_2$ column densities. Combining the results with the kinematic distances determined from the CO emission allows us to calculate physical properties such as bolometric luminosities and clump masses. 
To guarantee the consistency of our data with other studies, we compare the peak H$_2$ column densities obtained from dust continuum emission against the column density derived from the $^{12}$CO(2--1) emission. Although deviations of up to an order of magnitude are found for individual clumps, we find a good agreement for the general trend, and allot the deviations to local variations in the gas-to-dust ratio and CO-to-H$_2$ conversion factor. Furthermore, we compare the clump masses of our sample to a similar sample from \citep{Elia2013a}, finding the mean values to be almost identical, showing our methods to be reliable.
Our main findings are:
\begin{enumerate}[label={(\roman{enumi})}, leftmargin=0cm, itemindent=1.0cm,labelwidth=\itemindent,labelsep=0.1cm,itemsep=\baselineskip]
\item In general, we find the positions of the identified CO clouds to be strongly correlated with the highest column density parts of the \Hi emission. On the other hand, we were also able to identify a web of bridges, spurs and blobs of star forming regions spanning between the larger star forming regions, unveiling the complex three-dimensional structure of the outer Galaxy in unprecedented detail. Although the latter might be an indication of the outer Galaxy to be of a flucculent nature, a definite answer is difficult due to the influence of a large, expanding supershell (GSH242-03+37) in the survey area.
\item For the investigated clumps, we find the evolutionary stages to be well separated by the dust temperature, bolometric luminosity and luminosity-to-mass ratio, consistent with our results in \citep{Koenig2017}. However, we find the clump masses and peak column densities to be similar in the starless and protostellar phase, but find these quantities to be significantly higher for the YSO phase, indicating that the more massive clumps evolve significantly faster than their lower mass counterparts.
\item For the outer Galaxy we find only 24 sources with an inferred luminosity higher than that of an early B-Type star \citep[$\sim$$8$\,\msun][]{Mottram2011}, the masses of only 8 sources above the threshold where it would be likely to host a massive dense core or high-mass protostar according to \citet{Csengeri2014} or at most 40 sources above the threshold for high-mass star formation as determined by \citep{Kauffmann2010}. Even more so, we only find 2 methanol Class\,\textsc{ii} masers, and 10 known as well as 24 candidate \hii regions in the whole survey area of the outer Galaxy, indicating only a low fraction of star forming regions in the outer Galaxy to be able to form high-mass stars.
\item Investigating the influence of the expanding Galactic supershell GSH\,242--03+37 in detail, we find the physical properties and star formation activity of sources located within the walls to be not statistically different from sources located farther away. Nevertheless, as we have seen in Section\,\ref{sect:outer:properties:shell} we find the number-density of sources increased within the walls of the supershell, leading us to the conclusion that the expanding supershell supports the formation of clumps, but once they collapse has no further influence.
\end{enumerate}
In \papertwo we will use the sample of star forming regions characterized in the present work in order to compare the properties of the outer Galaxy with those found for the inner Galaxy and derive global trends for the Milky Way.
\begin{acknowledgements}
This research made use of Astropy\footnote{\url{http://www.astropy.org}}, a community-developed core Python package for Astronomy \citep{AstropyCollaboration2013} and the Astropy affilate software package photutils\footnote{\url{https://github.com/astropy/photutils}}.\\
This research made use of Montage, funded by the National Aeronautics and Space Administration's Earth Science Technology Office, Computation Technologies Project, under Cooperative Agreement Number NCC5-626 between NASA and the California Institute of Technology. Montage is maintained by the NASA/IPAC Infrared Science Archive.\\
This publication also makes use of data products from the Wide-field Infrared Survey Explorer, which is a joint project of the University of California, Los Angeles, and the Jet Propulsion Laboratory/California Institute of Technology, funded by the National Aeronautics and Space Administration.
\end{acknowledgements}
\bibliographystyle{aa}
\bibliography{paper}
\end{document}